# Rate Coefficients for Rotational State-to-State Transitions in $H_2O$ + $H_2$ Collisions as Predicted by Mixed Quantum/Classical Theory (MQCT)

Carolin Joy, Dulat Bostan, Bikramaditya Mandal, and Dmitri Babikov[*]

## ABSTRACT

*Aim:* A new database of collisional rate coefficients for transitions between the rotational states of $H_2O$ collided with $H_2$ background gas is developed. The goal is to expand over the other existing databases in terms of the rotational states of water (200 states are included here) and the rotational states of hydrogen (10 states). All four symmetries of ortho and para water combined with ortho, and para hydrogen are considered.

*Methods:* The mixed quantum/classical theory of inelastic scattering implemented in the code MQCT is employed. A detailed comparison with previous databases is conducted to ensure that this approximate method is sufficiently accurate. Integration over collision energies, summation over the final states of $H_2$ and averaging over the initial states of $H_2$ is carried out to provide state-to-state, effective, and thermal rate coefficients in a broad range of temperatures.

*Results:* The rate coefficients for collisions with highly excited $H_2$ molecules are presented for the first time. It is found that rate coefficients for rotational transitions in $H_2O$ molecules grow with the rotational excitation of $H_2$ projectiles and exceed those of the ground state $H_2$, roughly, by a factor of 2. These data enable more accurate description of water molecules in high-temperature environments, where the hydrogen molecules of background gas are rotationally excited, and the $H_2O$ + $H_2$ collision energy is high. The rate coefficients presented here are expected to be accurate up to the temperature of ~ 2000 K.

---

[*] Author to whom all correspondence should be addressed; electronic mail: dmitri.babikov@mu.edu
Chemistry Department, Wehr Chemistry Building, Marquette University, Milwaukee, Wisconsin 53201-1881, USA

# 1. INTRODUCTION

Water, often referred to as the essential source of life, is the most abundant molecule within biological systems and its probable extraterrestrial origin adds an intriguing dimension to its significance (Horneck & Baumstark-Khan 2012). Strong (maser-type) emissions from both gaseous and solid forms of water have been observed in many astrophysical environments such as planets, comets, interstellar clouds, and star forming regions (Ehrenfreund et al. 2003). The interstellar water vapor was discovered in 1969 in the Orion nebula by Charles Townes (Cheung et al. 1969). The discovery of water within the circumstellar disk of the V883 Orionis protostar, as observed by the Atacama Large Millimeter/submillimeter Array (ALMA), marks the first instance of water being inherited into a protoplanetary disk without significant changes to its composition (Tobin et al. 2023). Probing water in astrophysical media requires observations of lines that are thermally excited and do not display population inversion. Unfortunately, such rotational lines are blocked from ground-based observations due to the presence of water in Earth's atmosphere. Hence, space missions such as Infrared Space Observatory (ISO), the Submillimeter Wave Astronomy Satellite (SWAS), Odin, Hershel Space Observatory (HSO), and the recently launched mid-infrared space observatory James Webb Space Telescope (JWST) have played a pivotal role in advancing our understanding of water in space. In particular, high-excitation rotational lines of $H_2O$ are commonly detected in the youngest, most deeply embedded protostars, likely excited by far-UV irradiated shocks associated with outflows (Karska et al. 2014). Interpreting these observations requires numerical radiation transfer modeling and relies heavily on accurate collisional rate coefficients. In molecular clouds, the dominant collisional partner for $H_2O$ is $H_2$.

The collisional excitation of $H_2O$ by $H_2$ was in the focus of several theoretical studies over the past few decades. The first calculations of rate coefficients for this system were performed by Phillips, Maluendes & Green (1996), considering transitions between rotational levels up to $j = 3$ in both ortho- and para-$H_2O$ for temperatures ranging from 20 to 140 K.

Subsequent work by (Daniel, Dubernet & Grosjean 2011) employed an improved potential energy surface (PES) developed by (Valiron et al. 2008), made use of an accurate coupled-channel method of the MOLSCAT package (Hutson & Le Sueur 2019) and covered all four symmetries of the $H_2O + H_2$ system. They computed rate coefficients for transitions involving the first 45 rotational levels (up to $j = 11$) of $o$-$H_2O$ and $p$-$H_2O$, covering a broader temperature range of 5-1500 K. For $p$-$H_2O$ + $p$-$H_2$ symmetry they include rate coefficients for quenching of 44 rotationally excited states of water with rotational levels up to $j = 11$, collided with hydrogen in its ground rotational state, $j = 0$. Additionally, for collisions of $p$-$H_2O$ with $p$-$H_2$ in its excited rotational state $j = 2$ the rate coefficients are available for quenching of 19 excited states of water, up to $j = 7$. For $o$-$H_2O$ + $p$-$H_2$ symmetry, they have provided quenching rate coefficients for the excited 44 levels of water collided with both $j = 0$ and 2 of hydrogen, and additionally for the 9 excited states of ortho-water collided with hydrogen in $j = 4$ state. Overall, 8925 individual state-to-state transition rate coefficients are available from the work of Daniel et al (Daniel, Dubernet & Grosjean 2011).

In the recent work by (Żółtowski et al. 2021), the authors extended the calculations of $H_2O$ + $H_2$ rotational excitation rate coefficients to temperatures ranging from 10 to 2000 K and provided rate coefficients for 96 rotationally excited states of water with rotational levels up to $j = 17$, collided with hydrogen treated as a pseudo-atom. They found that to achieve convergence of the cross-sections for $H_2O$ energy levels up to $j = 17$, it was necessary to include energy levels up to $j = 29$ in the rotational basis. A reduced-dimensionality treatment of the $H_2$ molecule (that does not fully account for the rotational structure of the $H_2$ projectile) was adopted to make the scattering calculations computationally feasible. Moreover, to facilitate their calculations even further, they had to appeal to an approximate coupled-states (CS) treatment of scattering, also known as Coriolis-Sudden approximation (Pack 1984; Parker & Pack 1977). Overall, 9312 individual state-to-state transition rate coefficients are available from the work of Żółtowski et al. Only 1980 of these transitions overlap with the database of Daniel et al.

Recent quantum calculations by Stoecklin et al. (2019) (Garcia-Vázquez, Faure & Stoecklin 2024; Stoecklin et al. 2021) and (Wiesenfeld 2021, 2022) have investigated the coupling between rotation and bending of $H_2O$, albeit for a limited set of transitions and kinetic temperatures.

Despite of these efforts, there is still a lack of collisional rate coefficients at high temperatures and for highly excited rotational states of $H_2O$ and $H_2$. Indeed, the only collisional data available for highly excited rotational states of $H_2O$ is that of (Żółtowski et al. 2021) computed using an approximate treatment. To address this gap, it is desirable to compute rate coefficients for transitions involving a broader range of collision energies and higher levels of rotational excitation of both collision partners and using alternative methods. In this paper, we present the results of calculations using a rotational basis set that includes 100 states of each para-H$_2$O and ortho-H$_2$O (200 states total) and all rotational states of H$_2$ projectile up to $j = 10$, for collision energies up to 12000 cm$^{-1}$, without invoking the coupled-states approximation, i.e., retaining the physics of Coriolis coupling effect. This became possible due to recent developments of the mixed quantum/classical theory (Bostan et al. 2023, 2024; Joy et al. 2023; Mandal & Babikov 2023; Mandal et al. 2022; Mandal, Semenov & Babikov 2020) implemented in a user-ready massively parallel computer code MQCT [(Mandal et al. 2024; Semenov, Mandal & Babikov 2020)]. Many technical details of these calculations were reported in a recent paper (Joy et al. 2024). Here we describe how we computed rate coefficients, conduct a rigorous comparison of our results *vs.* those available in the literature, and make the database of new collisional rate coefficients for the H$_2$O + H$_2$ system available to the community.

## 2. DETAILS OF METHOD

In general, the state-to-state transition rate coefficients are obtained by averaging the corresponding cross sections $\sigma_{n_1 n_2 \to n'_1 n'_2}(U)$ over the Boltzmann-Maxwell distribution of kinetic energy $U$ at a certain temperature $T$. In the mixed quantum/classical calculations, this procedure has its specifics, that were discussed in detail in a recent paper (Mandal & Babikov 2023). The resultant formula is:

$$k_{n_1 n_2 \to n'_1 n'_2}(T) = \frac{v_{ave}(T)}{(k_B T)^2} \frac{e^{-\frac{\Delta E}{2k_B T}}}{(2j_1 + 1)(2j_2 + 1)}$$

$$\times \int_{U_{\min}}^{\infty} \tilde{\sigma}_{n_1 n_2 n'_1 n'_2}(U) e^{-\frac{U}{k_B T}\left[1 + \left(\frac{\Delta E}{4U}\right)^2\right]} \left[1 - \left(\frac{\Delta E}{4U}\right)^2\right] U dU \quad (1)$$

Here $n_1 n_2$ and $n'_1 n'_2$ indicate the initial and final states of two colliding molecules. The composite index $n_1 \equiv j_{1_{k_A k_C}}$ labels non-degenerate states of the first molecule (using a set of quantum numbers for an asymmetric top, such as water), whereas $n_2 \equiv j_2$ is used for the second molecule (since this is simply a linear rotor, H$_2$). The lower limit of integration in MQCT is $U_{\min} = |\Delta E|/4$ where $\Delta E = E_{n'_1 n'_2} - E_{n_1 n_2}$ is the energy difference between the final and initial states of the $n_1 n_2 \to n'_1 n'_2$ transition, which is negative for quenching and is positive for excitation processes, respectively. $v_{ave}(T)$ is the average collision speed, $k_B$ is the Boltzmann constant. Finally, $\tilde{\sigma}_{n_1 n_2, n'_1 n'_2}(U)$ is a weighted average of cross sections for excitation and quenching directions of the same transition, introduced to approximately satisfy the principle microscopic reversibility, as described elsewhere (Mandal & Babikov 2023):

$$\tilde{\sigma}_{n_1 n_2, n'_1 n'_2}(U) = \frac{1}{2}\{(2j_1 + 1)(2j_2 + 1)\sigma_{n_1 n_2 \to n'_1 n'_2}(U) \\ + (2j'_1 + 1)(2j'_2 + 1)\sigma_{n'_1 n'_2 \to n_1 n_2}(U)\} \qquad (2)$$

Our calculations were carried out for ten values of kinetic energy $U$ = 20.0, 41.28, 84.0, 170.47, 346.41, 703.89, 1430.0, 2906.3, 5906.0 and 12000 cm$^{-1}$. For numerical integration, a cubic spline of the entire integrant in Eq. (1) was constructed between these data points for each individual transition and was extrapolated towards the process threshold at low collision energies and towards the high collision energy limit using smooth analytic function as described elsewhere (Mandal & Babikov 2023). The result of this fitting was carefully checked for smoothness before using this dependence for numerical integration over the entire energy range.

Certain astrophysical applications may require the effective rate coefficients, $k^{n_2}_{n_1 \to n'_1}(T)$ which are obtained by summing state-to-state rate coefficients $k_{n_1 n_2 \to n'_1 n'_2}(T)$ over the final states $n'_2$ of H$_2$ for a given initial state $n_2$ of H$_2$:

$$k^{n_2}_{n_1 \to n'_1}(T) = \sum_{n'_2} k_{n_1 n_2 \to n'_1 n'_2}(T) \qquad (3)$$

The 'thermalized' state-to-state rate coefficient, $\bar{k}_{n_1 \to n'_1}(T)$ between the rotational states of the H$_2$O molecule can be obtained by averaging over the initial rotational levels $n_2$ of para- or ortho-H$_2$ as shown below:

$$\bar{k}^{\square}_{n_1 \to n'_1}(T) = \sum_{n_2} w_{n_2}\, k^{n_2}_{n_1 \to n'_1}(T) \qquad (4)$$

where the weights of initial states of the projectile H₂ can be expressed as: $w_{n_2} = (2j_2 + 1)\, e^{-\frac{E_2}{k_B T}}/Q_2(T)$ where the partition function for para- or ortho-hydrogen molecules, is $Q_{p/o}(T) = \sum_{n_2}(2j_2 + 1)\, e^{-\frac{E_2}{k_B T}}$.

## 2. RESULTS AND DISCUSSION

First, we present a comparison of our computed rate coefficients versus those from the databases of Żółtowski *et al* and of Daniel *et al* for a subset of 1980 transitions that are present in all three databases. For our results obtained with AT-MQCT method (Mandal et al. 2023; Mandal, Semenov & Babikov 2020) we will use abbreviation "AT". For the results of Żółtowski *et al* we will use abbreviation "CS" that corresponds to their coupled-states calculations. For the results of Daniel *et al* we will use abbreviation "CC" that corresponds to the coupled-channel method. We note that all three methods used the same potential energy surface of H₂O + H₂ system (Valiron et al. 2008).

A comparison of three sets of data is conveniently done using the method of analysis called a Dalitz plot (Babikov et al. 2002). Imagine that for a given transition the three methods give different rate coefficients $k_{AT}$, $k_{CS}$ and $k_{CC}$. We define three unitless variables (coordinates):

$$\zeta_{AT} = k_{AT}/(k_{AT} + k_{CS} + k_{CC}),$$

$$\zeta_{CS} = k_{CS}/(k_{AT} + k_{CS} + k_{CC}) \text{ and}$$

$$\zeta_{CC} = k_{CC}/(k_{AT} + k_{CS} + k_{CC}),$$

that satisfy the relation $\zeta_{AT} + \zeta_{CS} + \zeta_{CC} = 1$ and each vary in the range from zero to one. These variables are used to place one point (that corresponds to this transition) within the area of a triangular plot that has three axes crossing at 120 deg. Each point on a Dalitz plot corresponds to one transition and all datapoints fall within a triangle, as one can see in Figs. 1, 2 and 3. If a datapoint is close to the middle of the Dalitz plot, this means that we have $k_{AT} \sim k_{CS} \sim k_{CC}$ for this transition (rate coefficients from three databases are approximately the same). If a point is close to the edge of the triangle, this means that in one of databases the value of rate coefficient is much smaller than in the other two (*e.g.*, $k_{CS} \ll k_{AT} + k_{CC}$). If a point is in the corner of the triangle, this means that in one of databases the value of rate coefficient is much larger than in the other two (*e.g.*, $k_{CC} \gg k_{AT} + k_{CS}$). Finally, if the point is on the diagonal of the triangle and is off-center, it means that only two databases have similar rate coefficients (*e.g.*, $k_{AT} \sim k_{CC} \neq k_{CS}$).

In Fig. 1 we compare rate coefficients for 990 individual state-to-state transition rate coefficients $k_{n_1 n_2 \to n_1' n_2'}(T)$ for *p*-H$_2$O + *p*-H$_2$ collisions at four different temperatures: $T = 100$ K, 500 K, 1000 K, and 1500 K. A similar picture for 990 transitions in *o*-H$_2$O + *p*-H$_2$ is presented in the Supplemental Information (see Fig. S1). Note that all these transitions correspond to an elastic scattering of the ground state H$_2$ projectile, $(j_2 = 0) \to (j_2 = 0)$, since only these processes are included in the CS database of Żółtowski et al. From these figures we see that the results of all three methods are in relatively good agreement, because the majority of datapoints appear relatively close to the center of the Dalitz plot. One can notice that at lower temperature ($T = 100$ K) the distribution of datapoints spreads more along the CS axis and remains roughly symmetric around it, but spreads somewhat less along the other two directions. This means that the agreement is a bit better between the AT and CC databases, and these two sets show about the same differences when compared to the CS database. Still, at $T = 100$ K the center of distribution is in the middle of the plot, meaning that on average all three databases give similar rates of energy transfer. This picture, however, changes as the temperature rises. At $T = 1500$ K the center of distribution move off center, indicating that CS rate coefficients are somewhat higher than those from AT and CC databases. However, the distribution of datapoint becomes much more concentrated, which means that at higher temperature the agreement improves for most individual state-to-state transitions (with only a few outlying points). Figure S1 of Supplemental Information shows the same trends for *o*-H$_2$O.

In Fig. 2 we present the Dalitz plot analysis of the effective rate coefficients $k_{n_1 \to n_1'}^{n_2}(T)$ that include summation over the final states of the projectile H$_2$. For our database (AT) the sum over all final states up to $j_2 = 8$ was computed for all transitions, using Eq. (3). For CC data, the effective rate coefficients from Daniel *et al* were computed using their code and data retrieved from BASECOL database, which includes final states of H$_2$ ($j_2 = 0, 2, 4$). For the CS database, where H$_2$ is not allowed to be excited, state-to-state transition rate coefficients for collisions with H$_2$ ($j_2 = 0$) were used instead (*i.e.*, same as in Fig. 1). From Fig. 2 it becomes clear that the effect of rotational excitation of H$_2$ is quite significant. He majority of datapoints in this figure, except a few outliers, fall into 1/3 of the Dalitz plot area along one of its edges, indicating a better agreement between the AT and CC databases where the rotational excitation of $H_2$ is considered, and a much larger deviation of these two databases from the CS database, where $H_2$ is described as a pseudo-atom and the value of rate coefficient is considerably smaller. Amongst the four temperatures presented, this effect is more pronounced at $T = 100$ K, when the majority of datapoint are found very close to one of the edges (see Fig. 2), which means that the value of effective rate coefficients in AT and CC databases are much larger than those in the CS database. Figure S2 of Supplemental Information shows the same trends for $o$-$H_2O$.

Finally, in Fig. 3 we present the comparison of thermal rate coefficients $\bar{k}_{n_1 \to n_1'}(T)$ obtained by averaging over Boltzmann distribution of the initial states of $p$-$H_2$ projectiles. For our database (AT) the average over even initial states up to $j_2 = 8$ was computed for all transitions, using Eq. (4). For CC data, thermal rate coefficients from Daniel *et al* were computed using their code and data retrieved from BASECOL database. For the CS database, where $H_2$ was not allowed to be excited, state-to-state transition rate coefficients for collisions with $H_2(j_2 = 0)$ were used instead (*i.e.*, same as in Figs. 1 and 2). Interestingly, from Fig. 3 one concludes that those larger differences between the databases we saw at the level of effective rate coefficients (in Fig. 2), are largely removed at the level of thermal rate coefficients. In particular, in Fig. 3 we see that for the temperatures $T = 500, 1000$ and $1500$ K the distribution of datapoints returns closer to the center of Dalitz plot, which means the agreement is systematically better. On average, at $T = 500$ K the rates of CS database are slightly smaller than those from the other two databases, while at $T = 1500$ K they are slightly larger. At 1500 K the rates of our AT database are roughly in the middle between those from CC and CS databases, with CC giving the smallest rates, and CS giving the largest rates (on average). For the lowest temperature considered here, $T = 100$ K, the results of CS database are noticeably smaller than those from the other two databases. It should be noted, however, that the CS database was specifically built for high-temperature applications, and therefore this small disadvantage is not essential. Overall, it appears that, despite of approximations made during the calculations of CS database (the coupled-states approximation and the pseudo-atom treatment of $H_2$) this database provides thermal rate coefficients in good agreement with the other two databases, except may be at the temperatures $T \sim 100$ K and lower. Figure S3 of Supplemental Information shows the same trends for $o$-$H_2O$ + $p$-$H_2$.

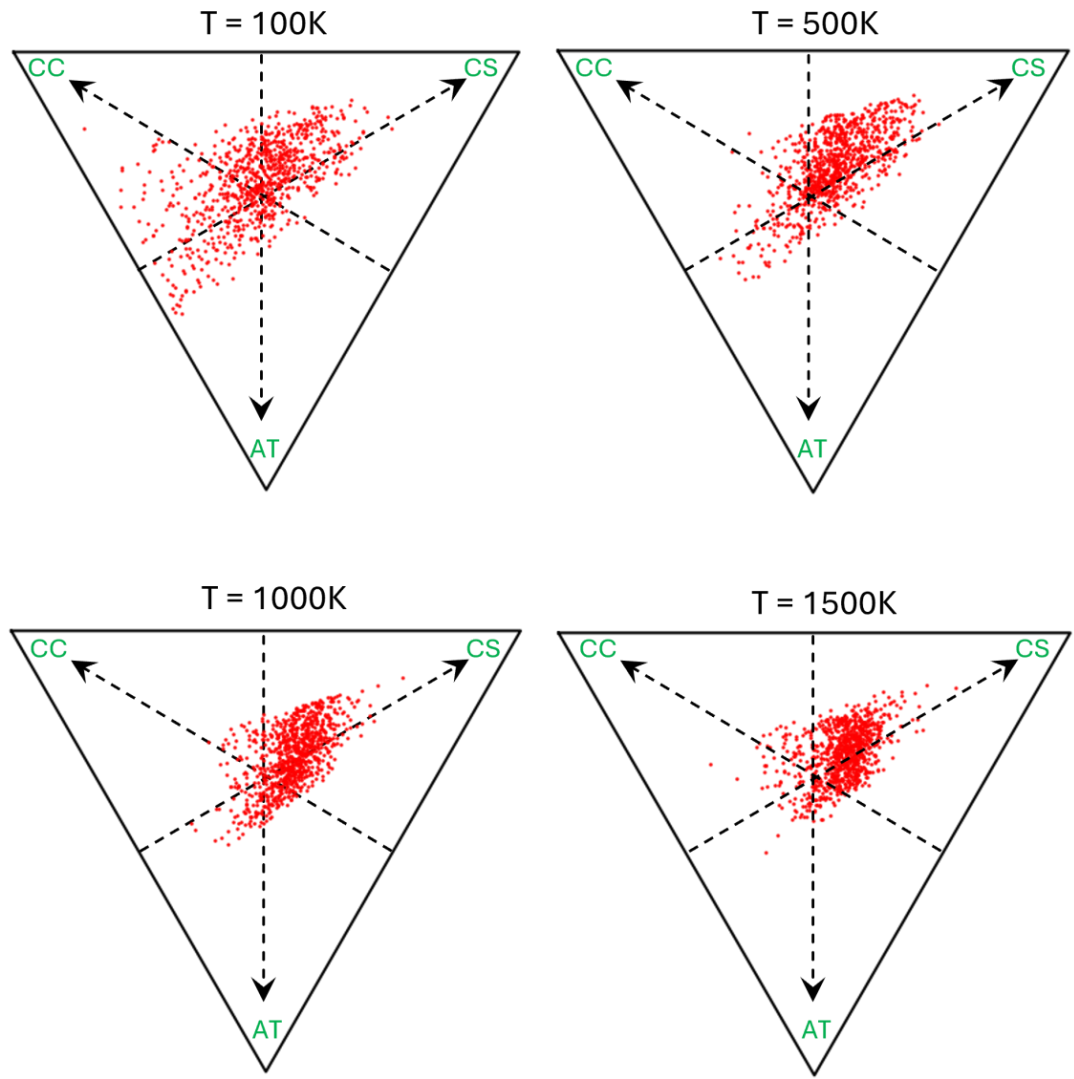

**Fig. 1:** Dalitz plots for the comparison of 990 *state-to-state* transition rate coefficients for *p*-H$_2$O + *p*-H$_2$ collisions at four different temperatures: 100 K, 500 K, 1000 K, and 1500 K. Each triangular plot represents a comparison between three computational methods: MQCT from this work (AT), full quantum close-coupling (CC) calculations (Daniel, Dubernet & Grosjean 2011), and full quantum coupled-states (CS) calculations (Faure et al. 2023) using MOLSCAT. Different red dots within each triangle represent different transitions.

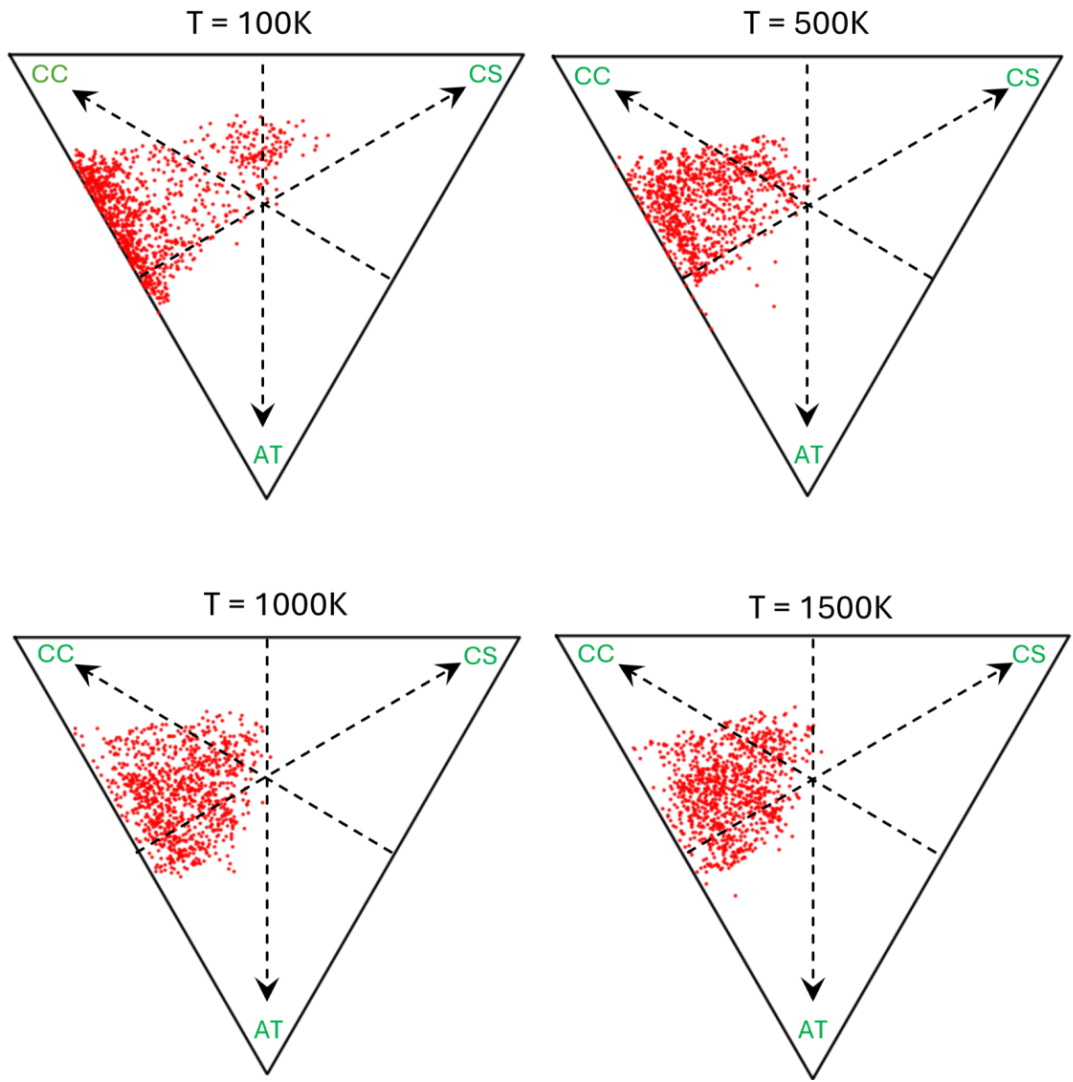

**Fig. 2:** Same as in Fig. 1 but for the *effective* rate coefficients defined by Eq. (3).

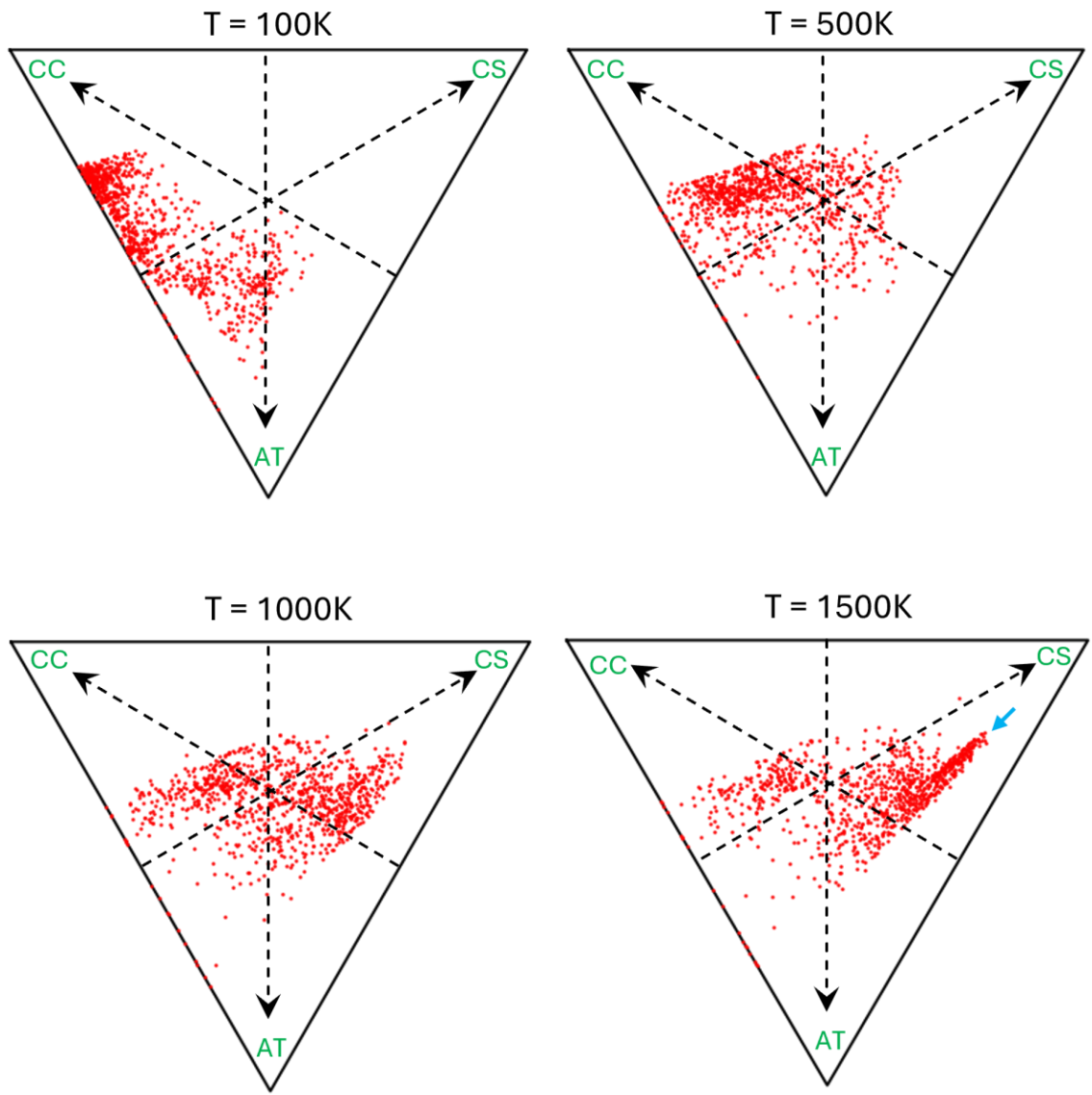

**Fig. 3:** Same as in Fig. 1 but for *thermal* rate coefficients defined by Eq. (4).

We should note, however, that looking at Fig. 3 one can notice a definite feature that becomes quite pronounced at 1500 K. It represents a group of points (indicated by blue arrow in the figure), where for many transitions the rate coefficients of CC database become much smaller compared to the other two databases. This is for *p*-$H_2O$, but if one looks at Fig. S3 for *o*-$H_2O$ this feature is not present. The distribution of points in Fig. S3 is close to the center of Dalitz plot, indicating good agreement between rate coefficients from all three databases, even at 1500 K.

We took a closer look at this difference and found that the code of Daniel *et al* retrieved from BASECOL database has an issue in the case of *p*-$H_2O$ + *p*-$H_2$. Namely, for the calculation of thermal rate coefficients Daniel *et al* proposed to use the effective rate coefficients computed for $H_2(j_2 = 2)$ as an educated guess for $H_2(j_2 = 4, 6, 8)$ that were not available, which is a very reasonable assumption that was implemented correctly in their code and works as expected for all transitions in *o*-$H_2O$ + *p*-$H_2$, as represented by Fig. 3S. However, for *p*-$H_2O$ + *p*-$H_2$ Daniel *et al* had the $H_2(j_2 = 2)$ data for some transitions, but not for all. For those cases they proposed to use the data they had for *p*-$H_2O$ + *o*-$H_2(j_2 = 1)$ as an approximation for *p*-$H_2O$ + *p*-$H_2(j_2 = 2, 4, 6, 8)$ which, again, could be a reasonable approximation. The problem is that this approximation described in their paper was not implemented in their code available at BASECOL, where the missing rate coefficients for the excited *p*-$H_2$ projectiles were basically set to be zero. As the result, thermal rate coefficients given by their code for *p*-$H_2O$ + *p*-$H_2$ are underestimated, and this becomes more severe at high temperature (as shown by blue arrow in Fig. 3). This finding does not downplay the quality of data and codes reported by Daniel *et al.* However, the users of their code downloaded from BASECOL should be aware of this and should implement the necessary modifications before using it. We found that this is needed only for *p*-$H_2O$ + *p*-$H_2$ symmetry while the other three symmetries were treated correctly.

Next, we conduct a detailed comparison of our AT database versus CS database using 9,312 transitions present in both databases. These include 4656 transitions for quenching of the lower 97 states of *p*-$H_2O$. In Figures 4, 5 and 6, we present a comparison of the individual state-to-state $k_{n_1 n_2 \to n_1' n_2'}(T)$, the effective $k_{n_1 \to n_1'}^{n_2}(T)$ and thermal $\bar{k}_{n_1 \to n_1'}(T)$ rate coefficients, respectively, in the two databases at four different temperatures: $T = 100, 500, 1000$ and $1500$ K. Similar pictures for 4656 transitions in *o*-$H_2O$ are presented in the Supplemental Information (see Figs. S4, S5 and S6). From Fig. 4 we see, first of all, that a systematically good agreement between the CS database (computed using an approximate quantum method) and our AT database (computed using a mixed quantum/classical theory approach) is found to persist through four orders of magnitude of rate coefficient values, which means that the agreement is good not only for the most intense transitions, but for all of them, including those transitions that are ~ 10,000 weaker. At higher temperatures $T = 500, 1000$ and $1500$ K the differences between rate coefficients for the *individual state-to-state* transitions in two databases (see Fig. 4) are within a factor of 2 for the majority of transitions. At the lower temperature $T = 100$ K many transitions show somewhat larger differences but, for the group of most intense transitions with rate coefficients larger than $10^{-11}$ cm$^3$s$^{-1}$, the differences between the two databases remain within a factor of 2. Importantly, for this group of transitions the differences between the two databases remain within a factor of 2 at all temperatures. In Table 1 we present the average difference between state-to-state rate coefficients in AT and CS databases at different temperatures. These values are always within ~ 50%. For para-H$_2$O the largest difference is observed at $T = 100$ K with AT rate coefficients being smaller, while for ortho-H$_2$O the larger difference is observed at higher temperatures with AT rate coefficients being larger.

From Fig. 5 for the *effective* rate coefficients, we see that the differences between the two databases become larger for many transitions, with AT rate coefficients being systematically larger (same property that we saw in Fig. 2 above), and particularly so at low temperature. However, in Fig. 5 one can notice that these larger differences concern mostly those transitions that are weaker. For the most intense transitions (with rate coefficients larger than $10^{-11}$ cm$^3$s$^{-1}$) the differences between the effective rate coefficients in the two databases are usually within a factor of 2, with only a few exceptions, and this is the case for all temperatures. The data for *o*-H$_2$O presented in Figs. S4 and S5 show similar trends.

**Table 1:** Average difference (AT - CS)/CS ×100% for the data presented in Fig. 4 for *p*-H$_2$O + *p*-H$_2$ and in Fig. S4 for *o*-H$_2$O + *p*-H$_2$ systems, at four different temperatures.

| $T$ (K) | Average difference (%) *p*-H$_2$O + *p*-H$_2$ | Average difference (%) *o*-H$_2$O + *p*-H$_2$ |
|---|---|---|
| 100 | -48.72 | -10.23 |
| 500 | 20.24 | 18.75 |
| 1000 | 21.30 | 44.57 |
| 1500 | 15.17 | 41.34 |

From Fig. 6 for *thermal* rate coefficients, we see a very good agreement between the two databases. Larger differences, present at the level of the effective rate coefficients, disappear at the level of thermal rate coefficients, due to averaging over the distribution of the initial states, which is the same effect we saw in Fig. 3 above. In fact, out of twelve frames presented in Figs. 4, 5 and 6, the best agreement is found in the last frame of Fig. 6, that corresponds to the comparison of thermal rate coefficients at high temperature, $T = 1500$ K. Here, for almost all transitions, the differences between rate coefficients in the two databases are within a factor of 2. However, it should be noted that at low temperature, $T = 100$ K (the left-most frame of Fig. 6), many transitions show much larger differences and deviate from the main trend that still follows the diagonal of the figure. For these transitions the rate coefficients of AT-MQCT database are larger than those of CS database. Also, in the left-most frame of Fig. 6 one can spot a group of points with large values of rate coefficients that exhibit differences exceeding the factor of 2. These transitions are: $7_{62} \rightarrow 6_{51}, 7_{71} \rightarrow 6_{60}, 8_{53} \rightarrow 7_{44}, 10_{19} \rightarrow 8_{17}, 8_{44} \rightarrow 7_{35}, 9_{28} \rightarrow 7_{26}, 8_{08} \rightarrow 7_{17}, 9_{19} \rightarrow 7_{17}, 9_{19} \rightarrow 8_{08}, 8_{17} \rightarrow 6_{15}, 8_{35} \rightarrow 6_{33}, 11_{1,11} \rightarrow 9_{19}$. Figure S6 for *o*-H$_2$O shows similar trends, with only a few transitions (among the intense once) exhibiting a difference over the factor of 2 between the two databases: $7_{70} \rightarrow 6_{61}, 11_{47} \rightarrow 11_{38}, 8_{54} \rightarrow 7_{43}, 8_{45} \rightarrow 6_{43}, 8_{36} \rightarrow 6_{34}, 9_{54} \rightarrow 8_{45}, 7_{25} \rightarrow 5_{23}, 9_{45} \rightarrow 8_{36}, 6_{61} \rightarrow 5_{50}, 11_{83} \rightarrow 10_{92}, 8_{27} \rightarrow 6_{25}, 7_{34} \rightarrow 5_{14}$.

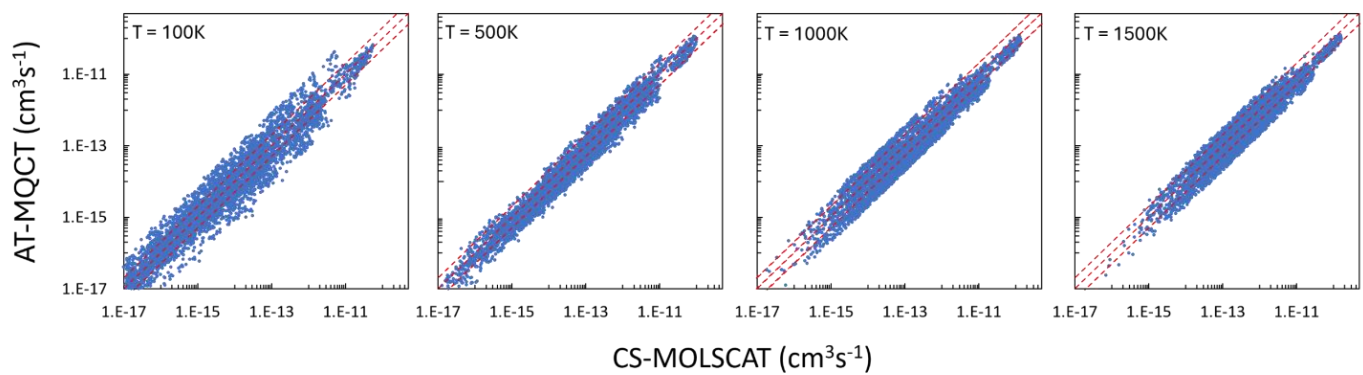

**Fig. 4:** Comparison of 4656 *state-to-state* transition rate coefficients for quenching of 97 lower states of *p*-$H_2O$ in collision with *p*-$H_2$ computed using AT-MQCT (this work) vs. those predicted by full quantum CS calculations using MOLSCAT (Faure et al. 2023) at four different temperatures: 100 K, 500 K, 1000 K, and 1500 K. Red dashed lines represent a factor of 2 difference.

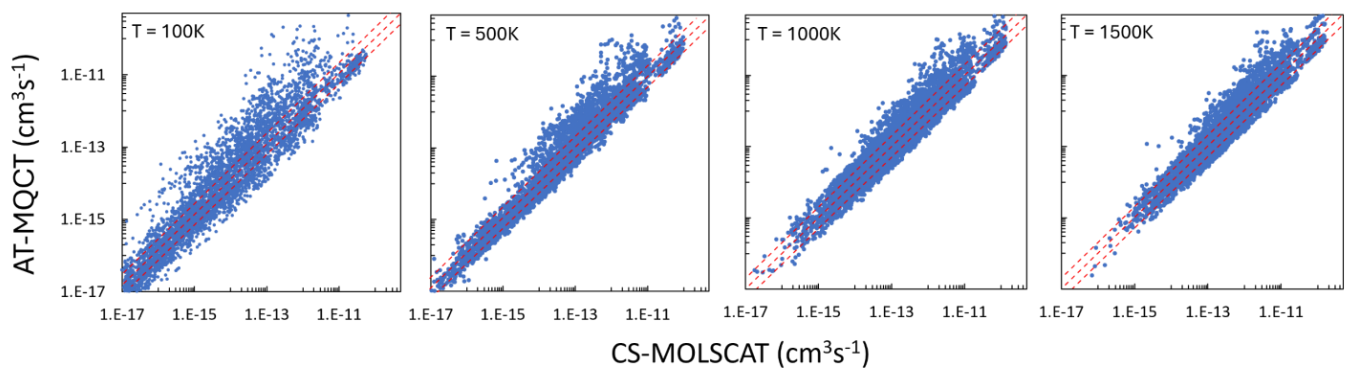

**Fig. 5:** Same as in Fig. 4 but for the *effective* rate coefficients defined by Eq. (3).

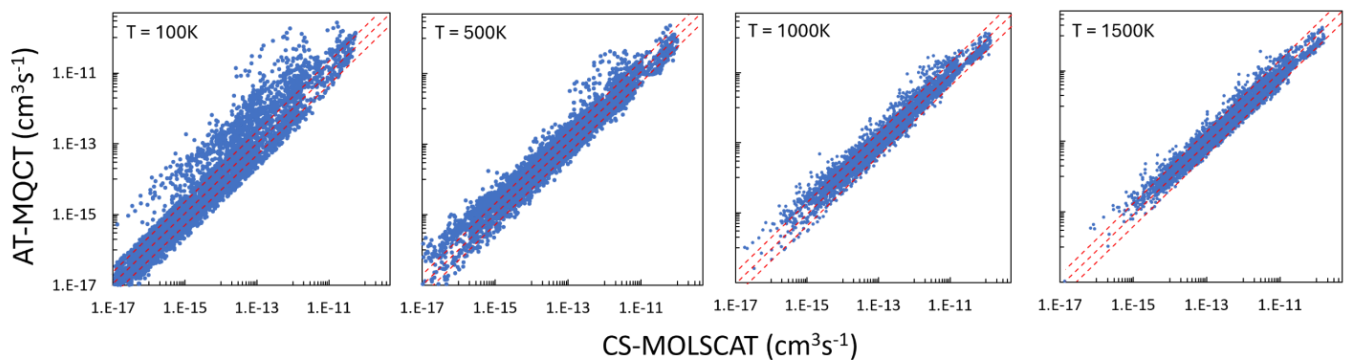

**Fig. 6:** Same as in Fig. 4 but for the *thermal* rate coefficients defined by Eq. (4).

While there is a generally good agreement between rate coefficients computed here and those available from literature, it is important to emphasize the advantage offered by our new database, which has an explicit dependence of rate coefficients on the rotational states of $H_2$ projectile in a broad range. In Fig. 7 we present thermal distribution of the rotational states of $H_2$ projectile at various temperatures. It demonstrates that an approximation in which the initial states of $H_2$ are restricted to the ground para-state ($j_2 = 0$) and ground ortho-state ($j_2 = 1$) is valid only at ~ 100 K or below, but even in this case it is important to consider ortho-$H_2$, because its population is significant, about 50% of para-$H_2$ population (purple curve in Fig. 7). At 500 K populations of $H_2$ states $j_2 = 2$ and 3 become significant (light-blue curve in Fig. 7). Importantly, at 1000 K the population of state $j_2 = 4$ exceeds that of $j_2 = 0$, and the population of $j_2 = 5$ is very close to that of $j_2 = 0$ (green curve in Fig. 7). At 2000 K populations of all states up to $j_2 = 7$ exceed that of $j_2 = 0$ and the population of $j_2 = 8$ is very close to that (red curve in Fig. 7). Therefore, the effect of these states should not be neglected.

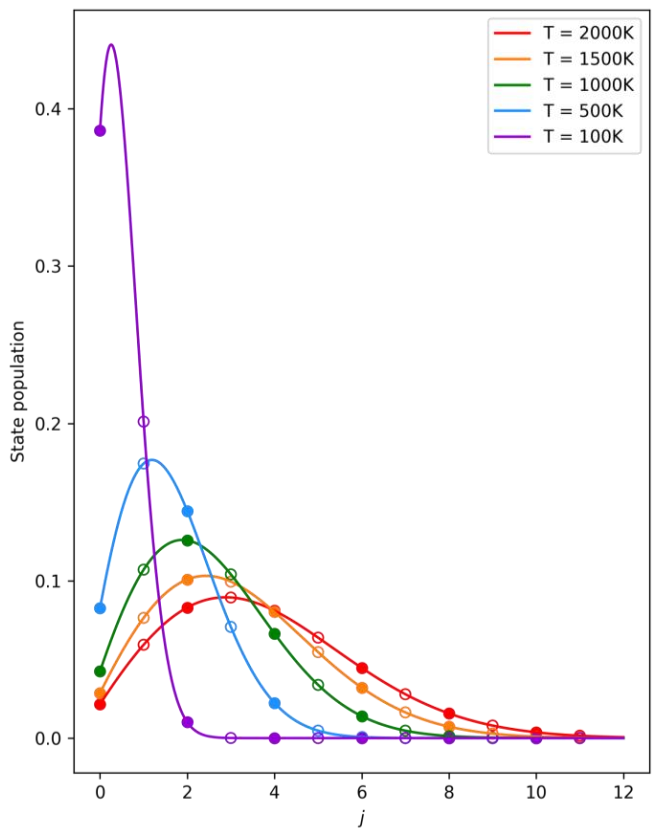

**Fig. 7:** Boltzmann distribution of $H_2$ rotational states as a function of rotational quantum number $j_2$ at five different temperatures between 100 and 2000 K, as shown by color in the figure. Para-$H_2$ states are represented by filled circles, while ortho-$H_2$ states are shown by empty circles.

This simple analysis shows that at higher temperatures the rotational states of H₂ up to $j_2 = 8$ must be included into consideration. Our database is the only one that offers this possibility and computes a true thermal average over the initial states of the projectile H₂ molecules, as appropriate for temperatures up to 2000 K. Moreover, using the effective rate coefficients $k_{n_1 \to n_1'}^{n_2}(T)$ that we provide for various initial states of H₂, users can compute an average over *any* distribution of the rotational states of the projectile (which may be different from the Boltzmann distribution). This can be necessary in non-LTE conditions, when the distribution of the initial states of H₂ is different from LTE, or when the rotational temperature of the background H₂ gas deviates from the kinetic temperature (Mandal & Babikov 2023). Our database offers this unique flexibility as well.

To illustrate these properties, we plotted in Figs. 8 and 9 the dependence of effective rate coefficients (for 20 most intense transitions in *p*-H₂O) on the initial state of H₂ projectile in the range $0 \leq j_2 \leq 9$, for several temperatures. Similar data for *o*-H₂O are presented in Supplemental Information, Figs. S7 and S8. In these figures, for convenience, we plotted *normalized* (or relative) values of the effective rate coefficients, computed as a unitless ratio between the rate coefficient for a certain value of $j_2$ of H₂ projectile and the same rate coefficient for the ground state of H₂ projectile. Namely, for para-H₂ we plotted $R_{j_2} = k_{n_1 \to n_1'}^{j_2} / k_{n_1 \to n_1'}^{j_2=0}$ at various values of temperature, while for ortho-H₂ we plotted $R_{j_2} = k_{n_1 \to n_1'}^{j_2} / k_{n_1 \to n_1'}^{j_2=1}$ for the same temperatures. These ratios can be viewed as scaling factors that one would have to apply to the effective rate coefficients computed for the ground state H₂ in order to obtain the effective rate coefficients for the exited states of H₂. Similar moieties were computed in (Daniel, Dubernet & Grosjean 2011) for a limited number of transitions with $j_2 = 2$ and 4. Our analysis is presented for many more transitions and up to $j_2 = 9$. For *p*-H₂O we included the following 20 transitions, identified using Fig. 4 as most intense at 1500 K: $1_{11} \to 0_{00}$, $2_{11} \to 2_{02}$, $2_{20} \to 1_{11}$, $2_{20} \to 2_{11}$, $3_{13} \to 2_{02}$, $3_{13} \to 2_{11}$, $3_{22} \to 3_{13}$, $3_{31} \to 2_{20}$, $3_{31} \to 3_{22}$, $4_{04} \to 3_{13}$, $4_{13} \to 4_{04}$, $4_{22} \to 4_{13}$, $4_{31} \to 4_{22}$, $5_{15} \to 4_{04}$, $5_{24} \to 4_{13}$, $5_{33} \to 4_{22}$, $5_{33} \to 5_{24}$, $5_{42} \to 4_{22}$, and $5_{42} \to 4_{31}$. For *o*-H₂O the 20 most intense transitions are listed in the Supplemental Information.

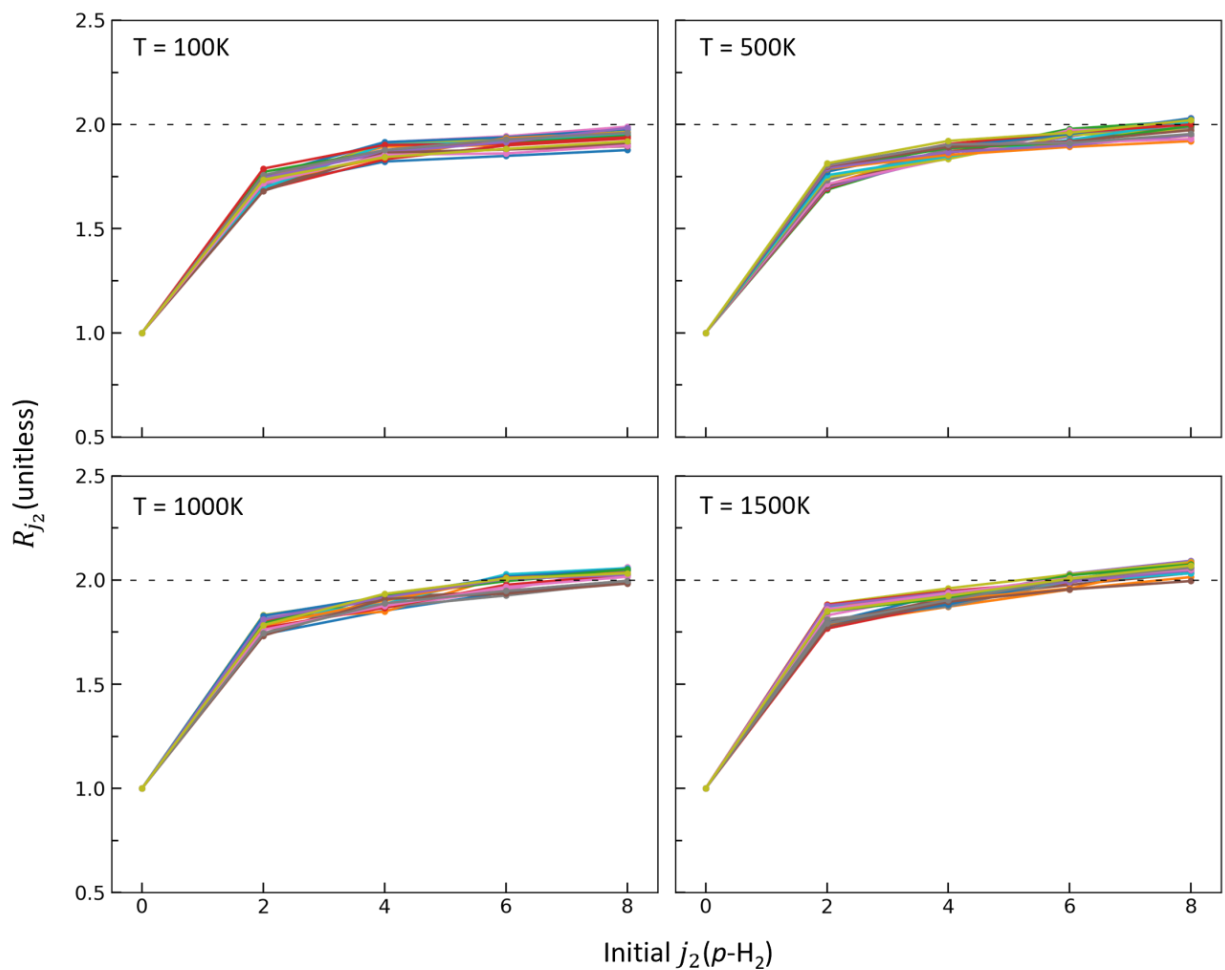

**Fig. 8:** The dependence of effective rate coefficients for *p*-H$_2$O on the initial state $j_2$ of *p*-H$_2$ projectile, computed by MQCT for 20 most intense transitions (identified using Fig. 4) at four temperatures as indicated in the figure.

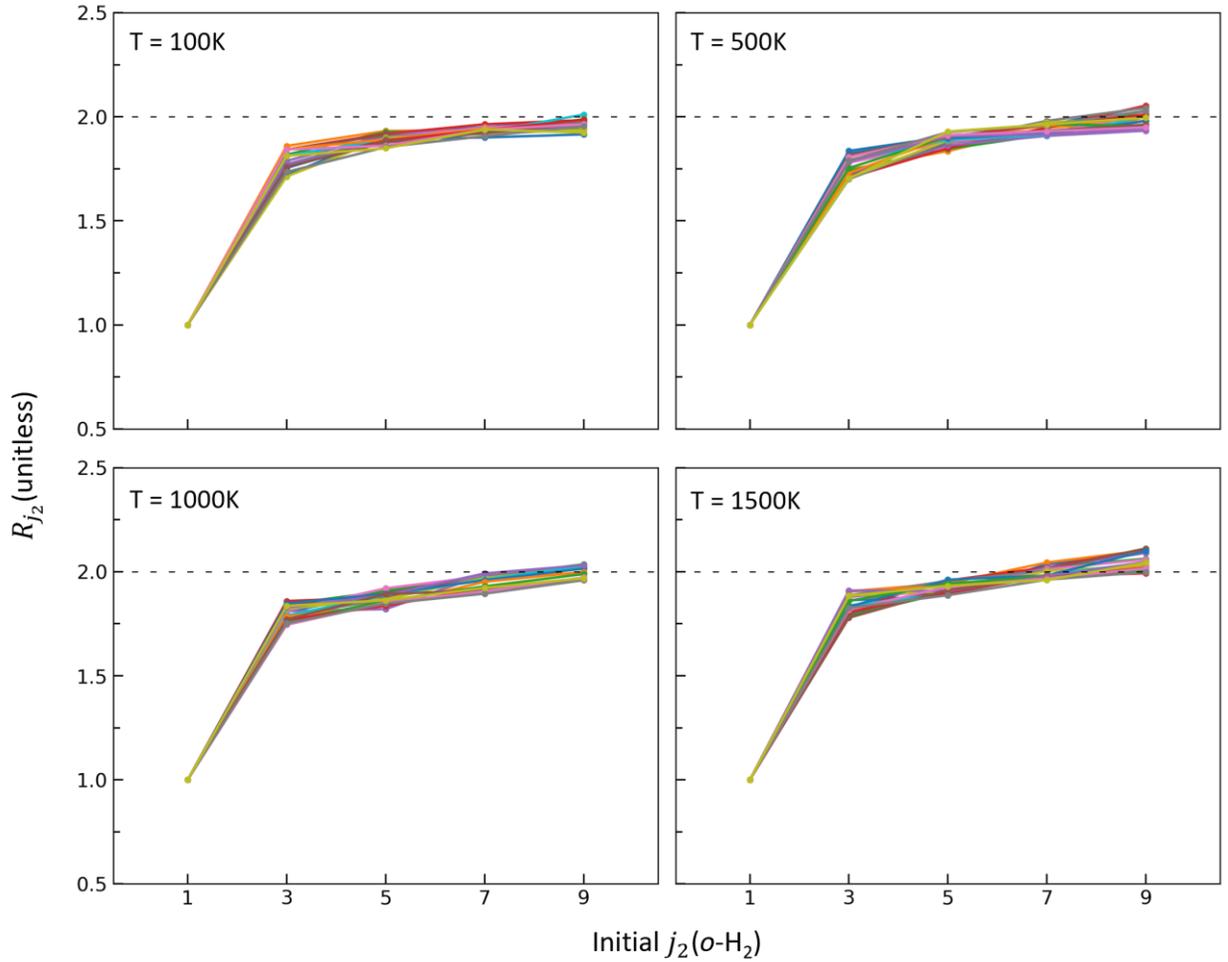

**Fig. 9:** Same as in Fig. 8 (for the same 20 transitions in *p*-H$_2$O) but as a dependence on the initial state $j_2$ of *o*-H$_2$ projectile.

From Figs. 8 one can see that the largest change of effective rate coefficients happens between the ground and first excited states of para-H$_2$ (i.e., as we move from $j_2 = 0$ to 2). This property was observed in the past by several authors who explored the effect of lower excited states of H$_2$ (Daniel, Dubernet & Grosjean 2011). However, it is often assumed that after that (at $j_2 = 4$ and above) the rate coefficients remain constant and equal to those with $j_2 = 2$. Our calculations show that this is not quite true. From Fig. 8 we see that rate coefficients keep growing when we look at $j_2 \geq 4$, although at a slower pace. From Fig. 9 we see that a similar trend is valid in the case of ortho-H$_2$. Namely, the largest increase of effective rate coefficients happens as we move from $j_2 =$ 1 to 3, but the rate coefficients keep growing in the range $j_2 \geq 5$ at a slower pace. Similar behavior is observed for $o$-$H_2O$ as illustrated by Figs. S7 and S8 in the Supplemental Information. On a semiquantitative level, we conclude that at higher temperatures (1500 K in Figs. 8, 9, S7 and S8), the values of effective rate coefficients for the highest states of $H_2$ we considered ($j_2 = 8$ and 9) are more than twice larger than those for the ground state $H_2$ ($j_2 = 0$ and 1), for many transitions. The value of $R_{j_2} = 2$ is shown in these figures by a dashed horizontal line, for convenience.

The dependencies of effective rate coefficients on the initial state of $H_2$ presented in Figs. 8, 9, S7 and S8 are all very similar and are relatively simple, but this simplicity is partially due to the way how these processes were selected (using Fig. 4 at 1500 K). However, there are different ways to select a small group of representative processes. Therefore, we also tried to identify 20 most important processes at each individual temperature and for each individual symmetry (of $p$- or $o$-$H_2O$ collided with $p$- or $o$-$H_2$) using the values of effective rate coefficients (like those from Fig. 5, rather than Fig. 4) that include summation over the final states of $H_2$ and therefore are more directly related to the final values of thermally averaged rate coefficients. These data are presented in Figs. S9, 10, 11 and 12 of Supplemental Information. Overall, they demonstrate similar dependencies with rate coefficients growing roughly by a factor of 2 when the rotational quantum number of $H_2$ is raised from $j_2 = 0$ and 1 to $j_2 = 8$ and 9, but these dependencies are more complex and show much more variations between different transitions in $H_2O$, different temperatures and different symmetries. Therefore, the use of actual data computed in this work, rather than a simplified scaling law that can be easily deduced from Figs. 8, 9, S7 and S8, is expected to give more accurate results and therefore is recommended.

## 3. CONCLUSION

In this paper we presented a new database of state-to-state, effective, and thermal rate coefficients for transitions between the rotational states of $H_2O$ collided with $H_2$ background gas. All four symmetries of ortho- and para-water combined with ortho- and para-hydrogen were considered. This database offers a significant expansion over the other existing databases in terms of the rotational states of water (200 states) and the rotational states of hydrogen (10 states). A detailed comparison with previous databases is presented which demonstrates that the approximate mixed quantum/classical theory employed in this work (MQCT) is sufficiently accurate. The behavior of rate coefficients for collisions with highly excited $H_2$ molecules is presented for the first time. It shows that collisional rate coefficients for rotational transitions in $H_2O$ molecules grow with the rotational excitation of $H_2$ projectiles and exceed those of the ground state $H_2$, roughly, by a factor of 2. These findings are important for accurate description of water molecules in high-temperature environments where the hydrogen molecules of background gas are rotationally excited, and the collision energy is high. The rate coefficients presented here are expected to be accurate up to the temperature of ~ 2000 K, when the vibrational excitation of $H_2O$ bending mode (not considered here) will likely start playing some role.

The individual state-to-state transition rate coefficients computed in this work will be submitted to the BASECOL database. Our codes that generate effective and thermal rate coefficients are available from the GitHub site: https://github.com/MarquetteQuantum/MOLRATES. The effective rate coefficients are available for a broad range of $H_2$ rotational states and can be employed for the modeling of collisional energy transfer in non-equilibrium conditions, when the distribution of the rotational states of $H_2$ deviates from the Boltzmann distribution at given kinetic temperature.

**CONFLICT OF INTEREST**

There are no conflicts to declare.

**ACKNOWLEDGMENTS**

This research was supported by NASA, grant number 80NSSC24K0208. D. Babikov acknowledges the support of the Way Klingler Research Fellowship and the Haberman-Pfletshinger Research Fund. CJ acknowledges the support of the Schmitt Fellowship. D. Bostan acknowledges the support of the Eisch Fellowship and the Bournique Memorial Fellowship. We used HPC resources at Marquette funded in part by the National Science Foundation award CNS-1828649.

# "Rate Coefficients for rotational state-to-state transitions in $H_2O + H_2$ collisions as predicted by Mixed Quantum/Classical Theory (MQCT)"

by Carolin Joy, Dulat Bostan, Bikramaditya Mandal, and Dmitri Babikov [*]

*Chemistry Department, Marquette University, Milwaukee, Wisconsin 53201-1881, USA*

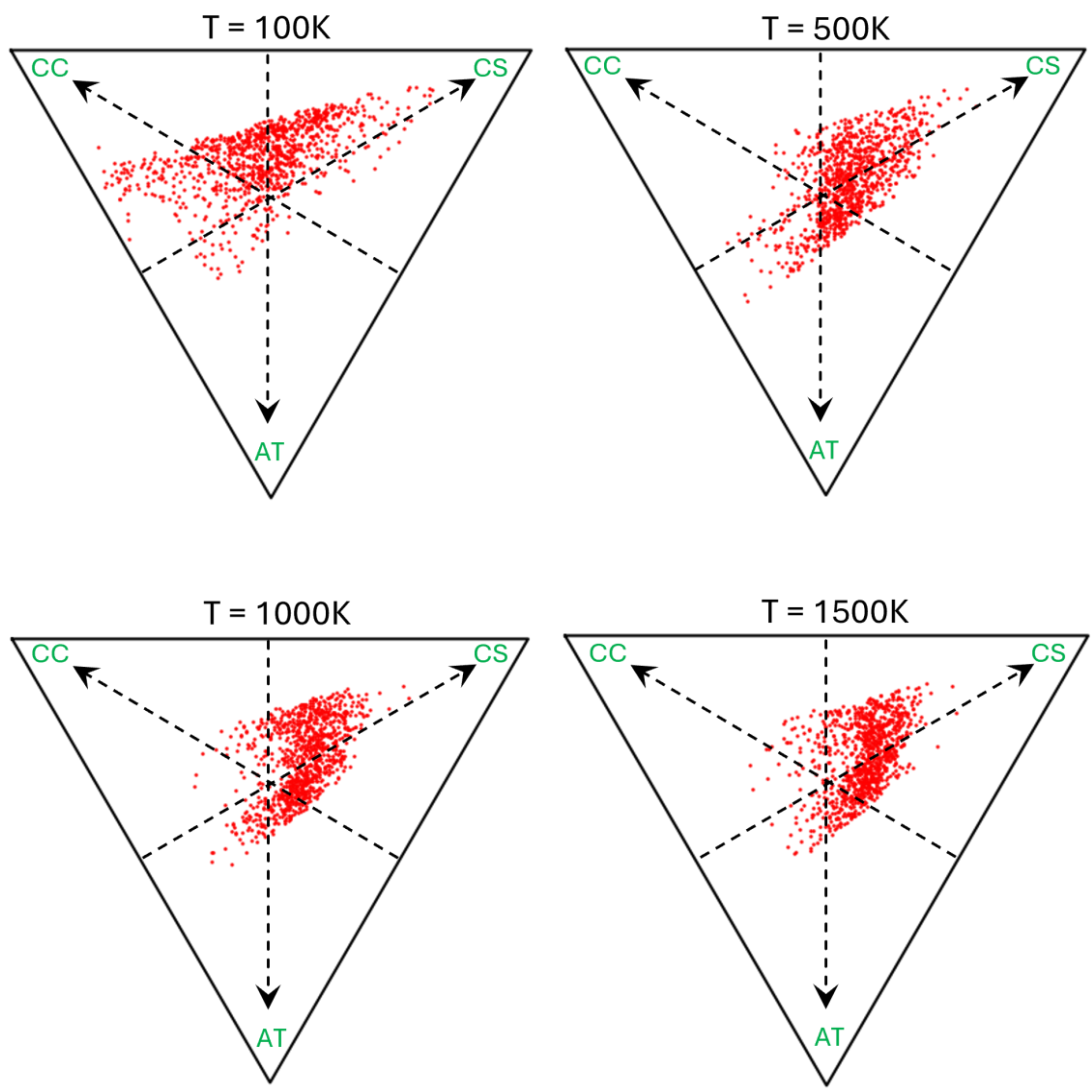

**Fig. S1:** Dalitz plots for the comparison of 990 *state-to-state* transition rate coefficients for *o*-$H_2O$ + *p*-$H_2$ collisions at four different temperatures: 100K, 500K, 1000K, and 1500K. Each triangular plot represents a comparison between three computational methods: MQCT from this work (AT), full quantum close-coupling (CC) calculations (Daniel et al., 2011), and full quantum coupled-states (CS) calculations (Faure et al., 2023) using MOLSCAT. Different red dots within each triangle represent different transitions.

---

[*] Author to whom all correspondence should be addressed; electronic mail: dmitri.babikov@mu.edu

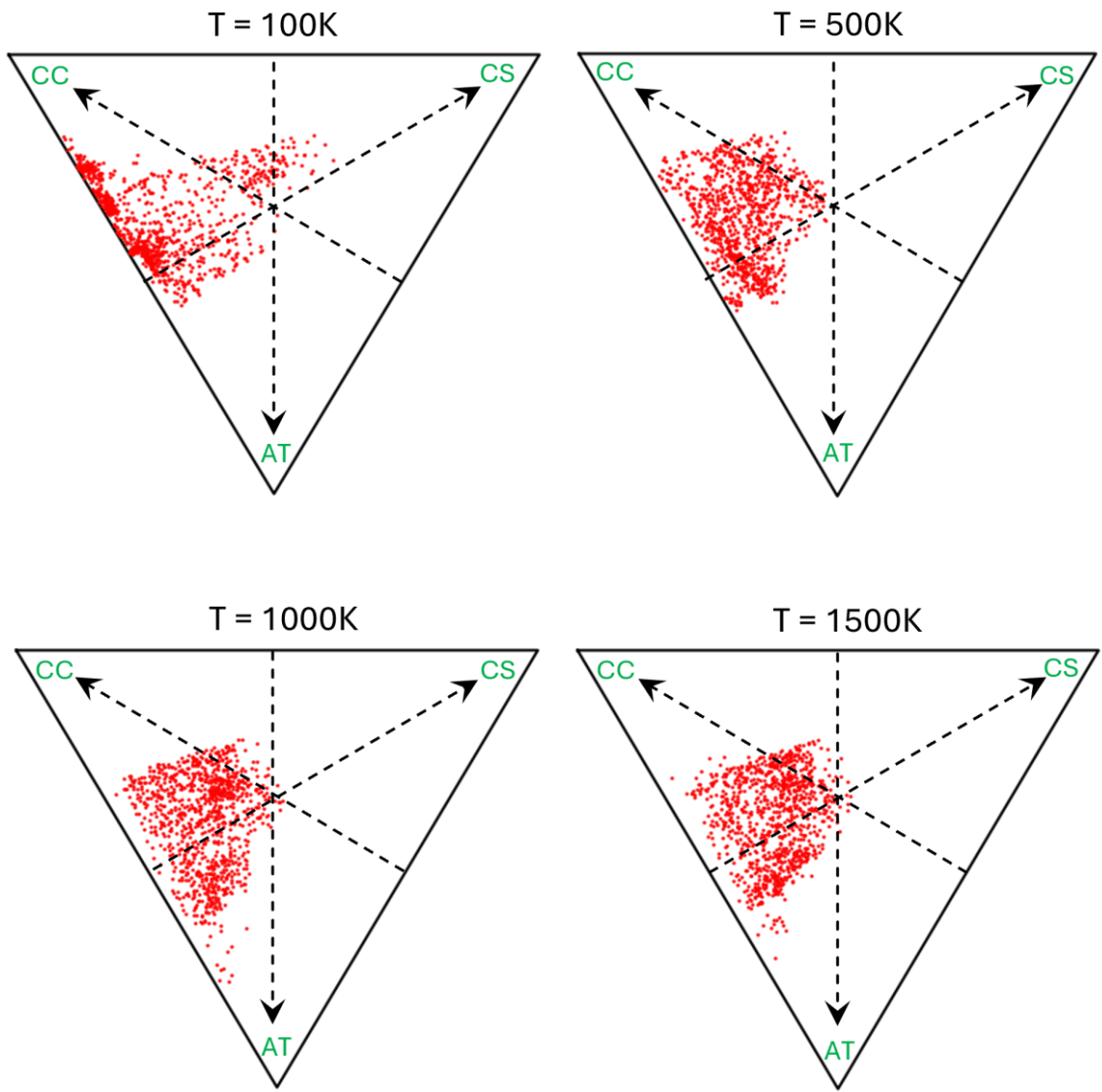

**Fig. S2:** Same as in Fig. S1 but for the *effective* rate coefficients defined by Eq. (3) in the main text.

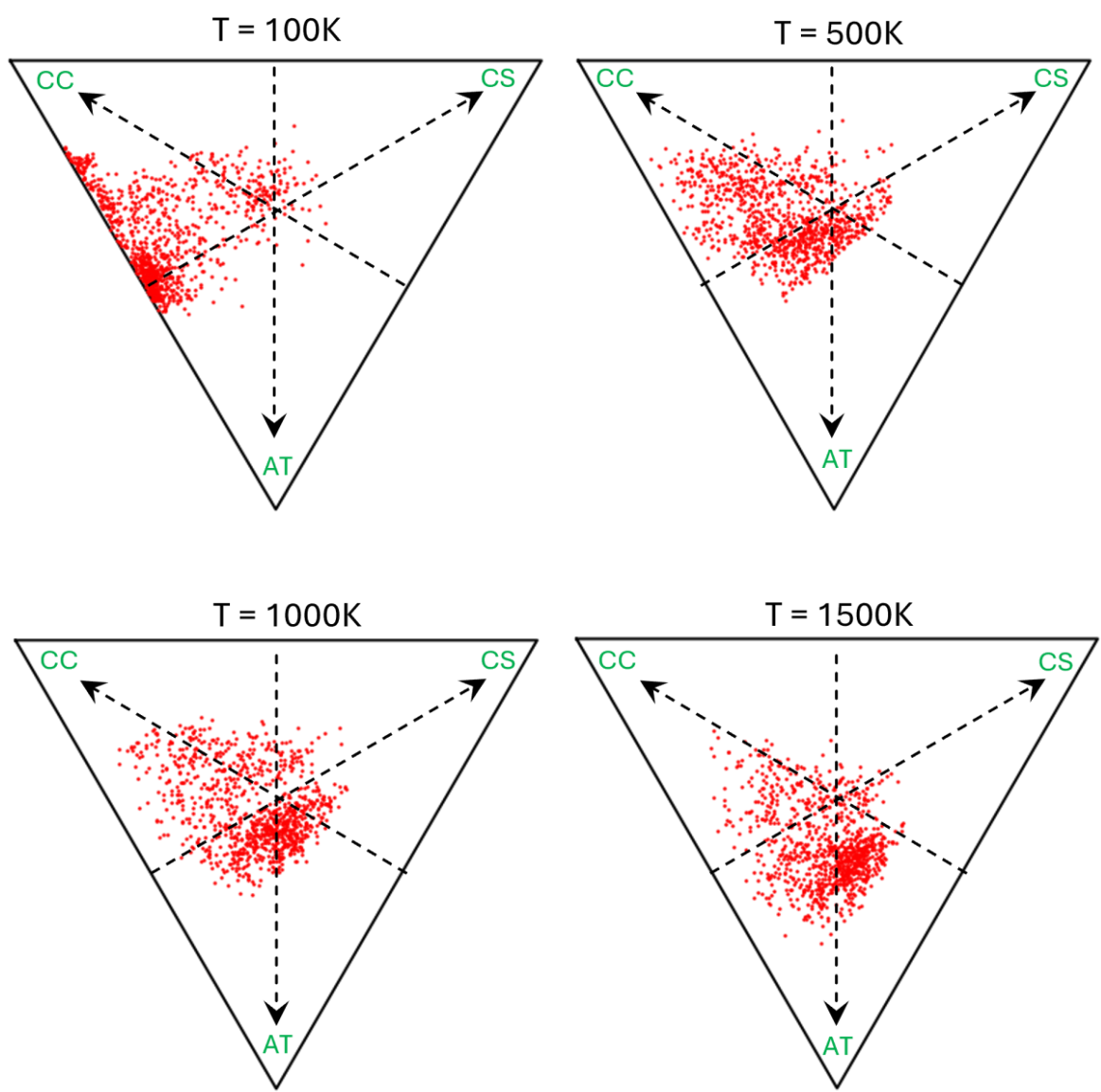

**Fig. S3:** Same as in Fig. S1 but for *thermal* rate coefficients defined by Eq. (4) in the main text.

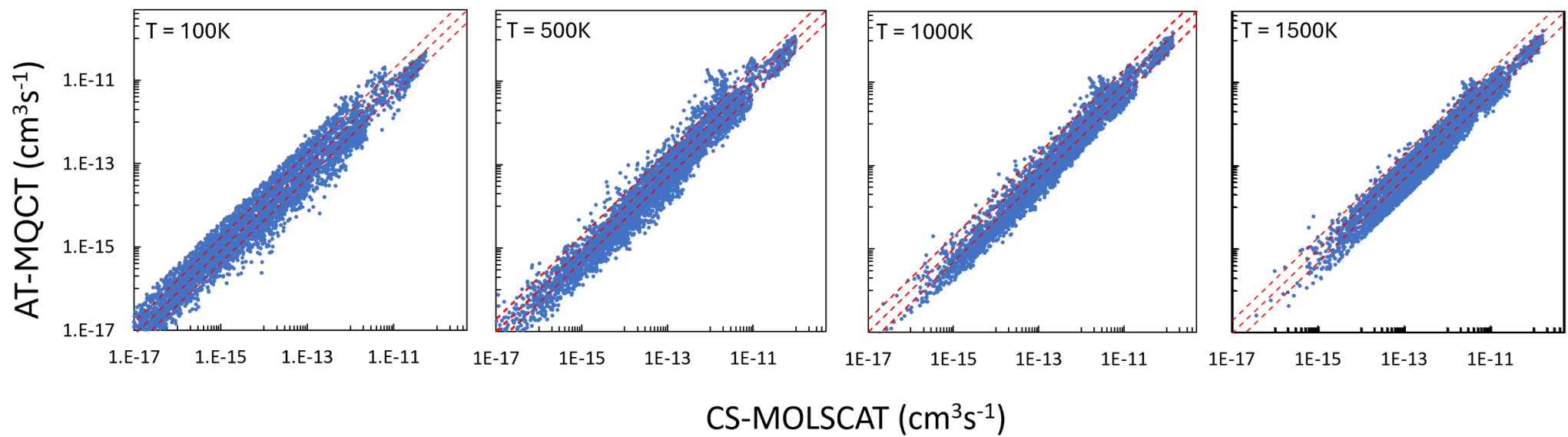

**Fig. S4:** Comparison of 4656 *state-to-state* transition rate coefficients for quenching of 97 lower states of *o*-$H_2O$ in collision with *p*-$H_2$ computed using AT-MQCT (this work) vs. those predicted by full quantum CS calculations using MOLSCAT (Faure et al., 2023) at four different temperatures: 100 K, 500 K, 1000 K, and 1500 K. Red dashed lines represent a factor of 2 difference.

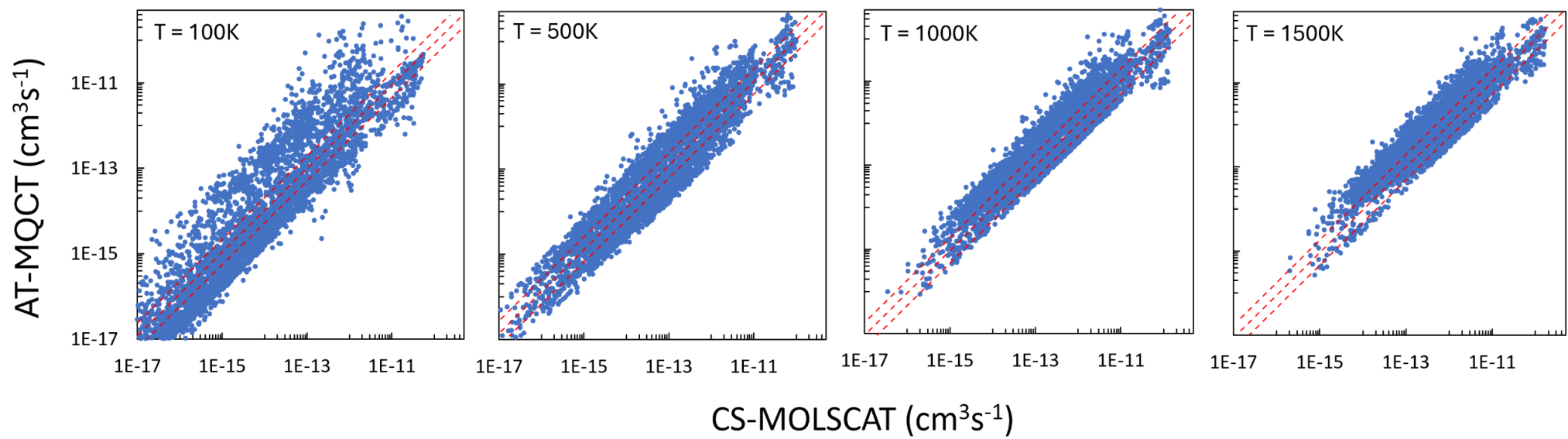

**Fig. S5:** Same as in Fig. S4 but for the *effective* rate coefficients defined by Eq. (3) in the main text.

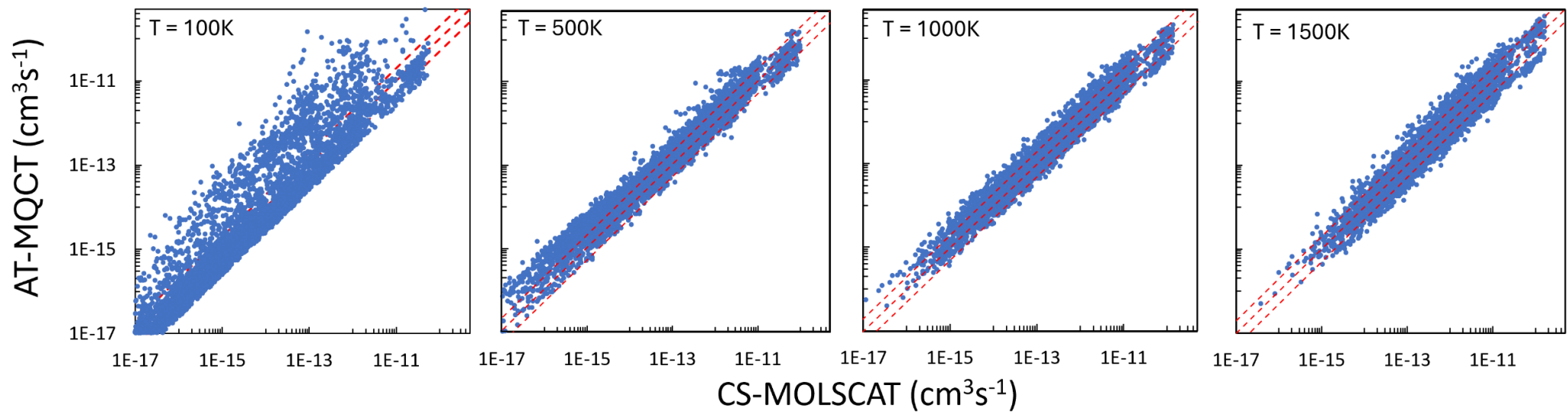

**Fig. S6:** Same as in Fig. S4 but for the *thermal* rate coefficients defined by Eq. (4) in the main text.

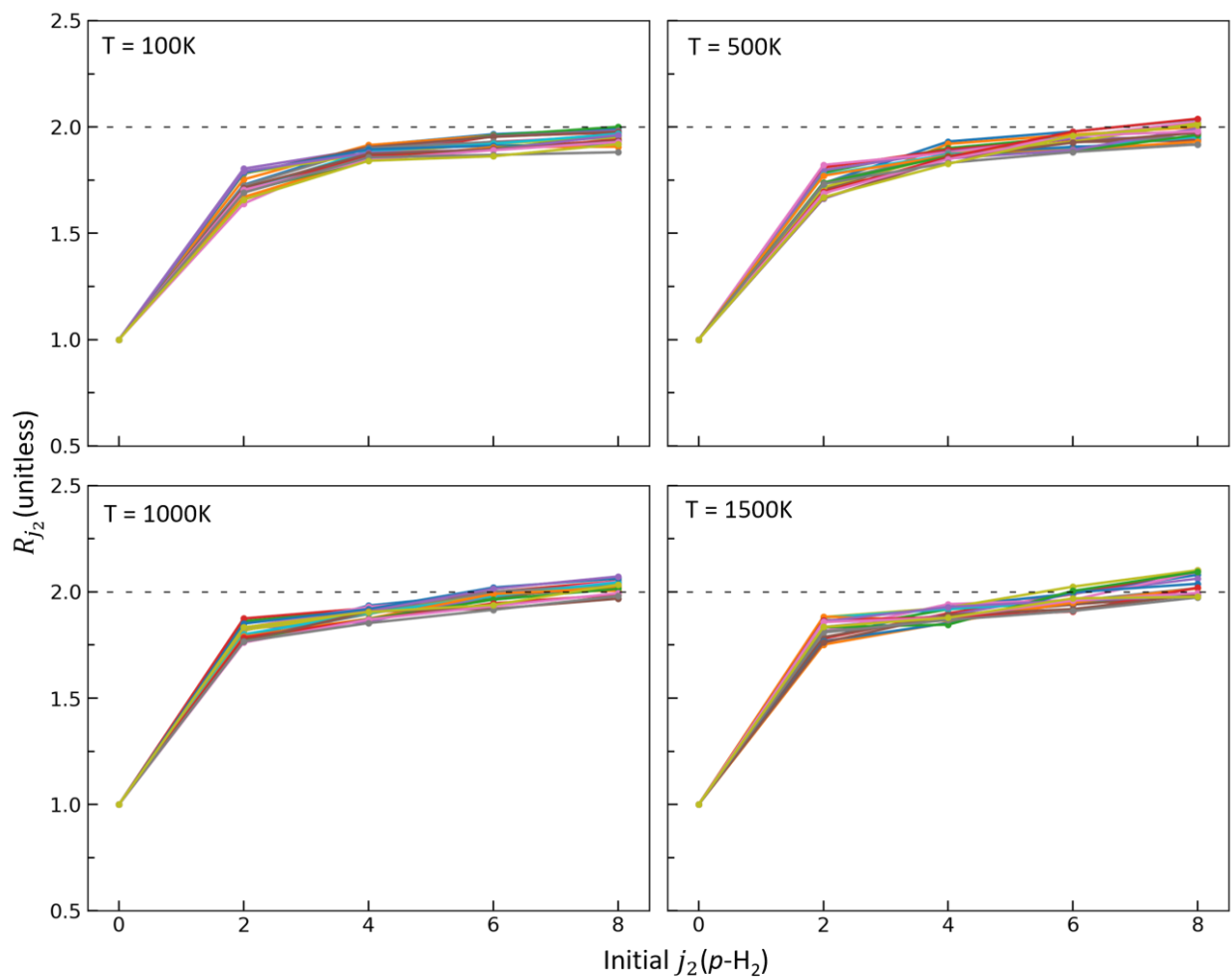

**Fig. S7:** The dependence of effective rate coefficients for $o$-H$_2$O on the initial state $j_2$ of $p$-H$_2$ projectile, computed by MQCT for 20 most intense transitions (identified using Fig. S4 at 1500 K) at four temperatures as indicated in the figure.

The transitions in $o$-H$_2$O are: $1_{10} \to 1_{01}$, $2_{12} \to 1_{01}$, $2_{21} \to 1_{10}$, $2_{21} \to 2_{12}$, $3_{03} \to 2_{12}$, $3_{12} \to 3_{03}$, $3_{21} \to 3_{12}$, $3_{30} \to 2_{21}$, $4_{14} \to 3_{03}$, $4_{23} \to 3_{12}$, $4_{23} \to 4_{14}$, $4_{32} \to 3_{21}$, $4_{41} \to 3_{30}$, $5_{05} \to 4_{14}$, $5_{14} \to 5_{05}$, $5_{23} \to 5_{14}$, $5_{32} \to 4_{23}$, $5_{32} \to 5_{23}$, $5_{41} \to 4_{32}$, and $5_{50} \to 4_{41}$.

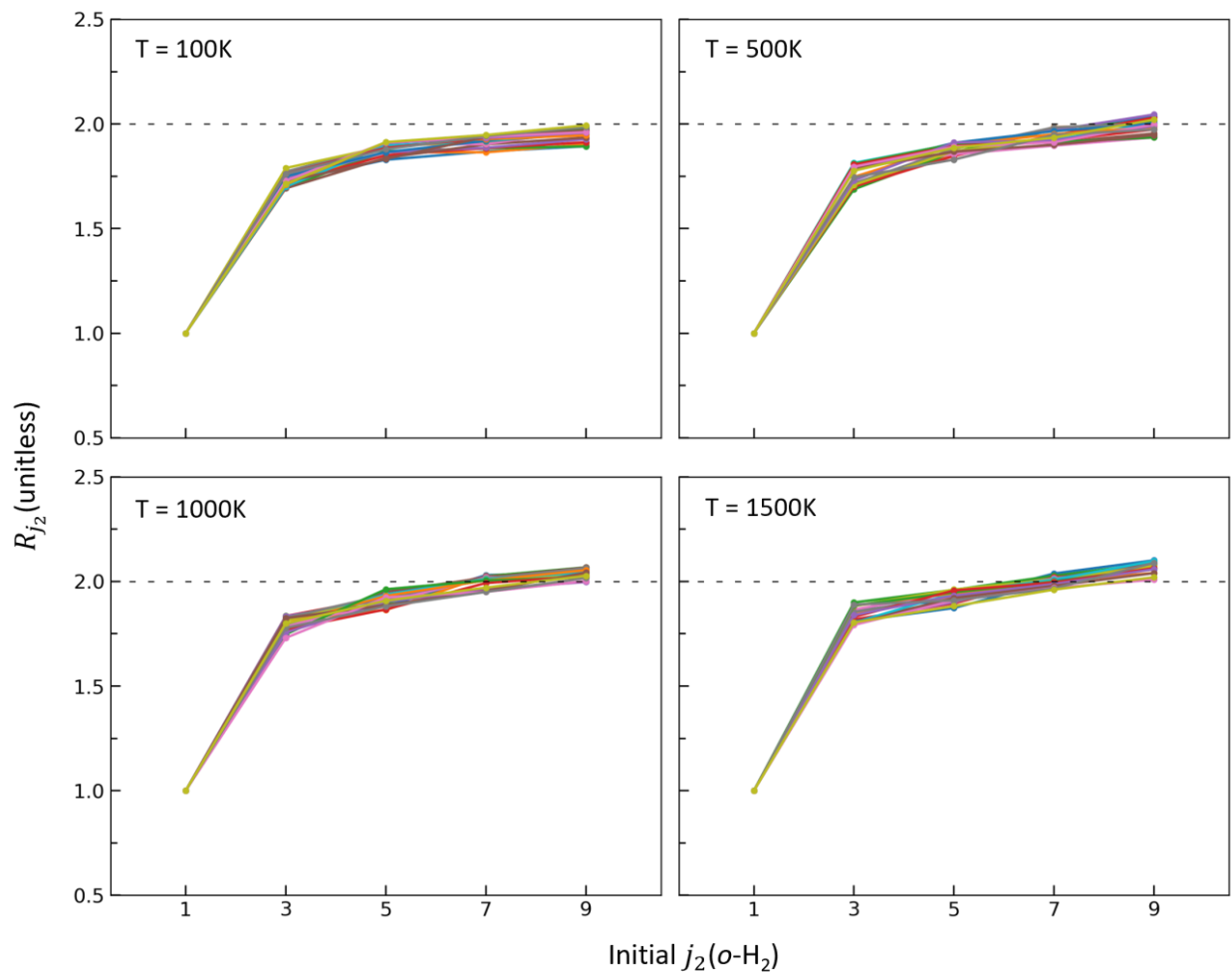

**Fig. S8:** Same as in Fig. S7 (for the same 20 transitions in *o*-H₂O) but as a dependence on the initial state $j_2$ of *o*-H$_2$ projectile.

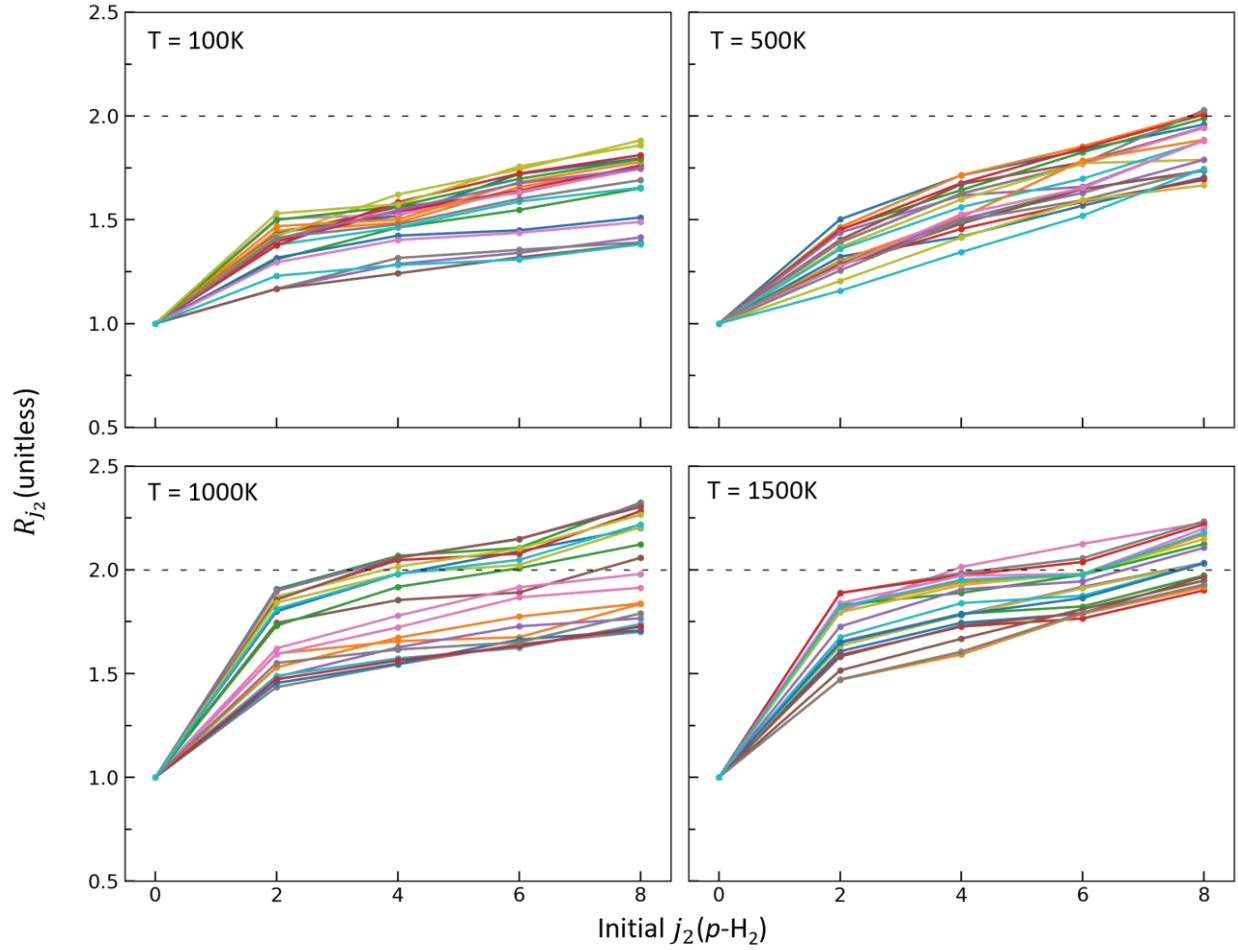

**Fig. S9:** The dependence of effective rate coefficients for $p$-H$_2$O on the initial state $j_2$ of $p$-H$_2$ projectile, computed by MQCT for 20 most intense transitions identified using Fig. 5 for each individual temperature (indicated in the figure). The transitions are:

For T = 100K: $10_{010} \to 9_{19}$, $12_{57} \to 7_{17}$, $12_{48} \to 8_{08}$, $10_{55} \to 9_{64}$, $10_{46} \to 10_{37}$, $9_{46} \to 8_{53}$, $9_{19} \to 8_{08}$, $8_{62} \to 8_{53}$, $8_{53} \to 7_{44}$, $8_{44} \to 8_{35}$, $8_{44} \to 7_{35}$, $8_{44} \to 6_{42}$, $8_{26} \to 6_{24}$, $8_{08} \to 7_{17}$, $8_{08} \to 5_{15}$, $7_{71} \to 6_{60}$, $7_{62} \to 6_{51}$, $7_{53} \to 6_{42}$, $7_{44} \to 6_{33}$, and $7_{17} \to 6_{06}$.

For T = 500K: $10_{010} \to 9_{19}$, $12_{57} \to 12_{148}$, $10_{46} \to 10_{37}$, $9_{19} \to 8_{08}$, $8_{53} \to 7_{53}$, $8_{53} \to 7_{44}$, $8_{44} \to 7_{35}$, $8_{44} \to 6_{42}$, $8_{26} \to 6_{24}$, $8_{08} \to 7_{17}$, $7_{71} \to 6_{60}$, $7_{62} \to 6_{51}$, $7_{53} \to 6_{42}$, $7_{44} \to 6_{33}$, $7_{26} \to 5_{24}$, $7_{17} \to 6_{06}$, $6_{60} \to 5_{51}$, $6_{51} \to 5_{42}$, $6_{06} \to 5_{15}$, and $6_{42} \to 5_{33}$.

For T = 1000K: $10_{010} \to 9_{19}$, $12_{57} \to 12_{148}$, $8_{53} \to 7_{53}$, $8_{53} \to 7_{44}$, $8_{44} \to 7_{35}$, $8_{44} \to 7_{44}$, $8_{35} \to 6_{33}$, $8_{17} \to 7_{26}$, $8_{08} \to 7_{17}$, $7_{71} \to 6_{60}$, $7_{71} \to 5_{51}$, $7_{62} \to 6_{51}$, $7_{53} \to 7_{44}$, $7_{53} \to 6_{42}$, $7_{44} \to 7_{35}$, $7_{44} \to 6_{33}$, $7_{35} \to 6_{24}$, $7_{26} \to 6_{15}$, $7_{17} \to 6_{06}$, and $6_{60} \to 5_{51}$.

For T = 1500K: $11_{111} \to 10_{010}$, $11_{111} \to 9_{19}$, $10_{010} \to 9_{19}$, $10_{010} \to 8_{08}$, $10_{19} \to 9_{28}$ $10_{19} \to 8_{17}$, $9_{46} \to 8_{35}$, $9_{37} \to 8_{26}$, $9_{28} \to 8_{17}$, $9_{19} \to 8_{08}$, $8_{53} \to 7_{44}$, $8_{44} \to 7_{44}$, $8_{26} \to 8_{17}$, $8_{17} \to 7_{26}$, $8_{08} \to 7_{17}$ $7_{71} \to 6_{60}$, $7_{62} \to 6_{51}$, $7_{53} \to 6_{51}$, and $7_{53} \to 6_{42}$.

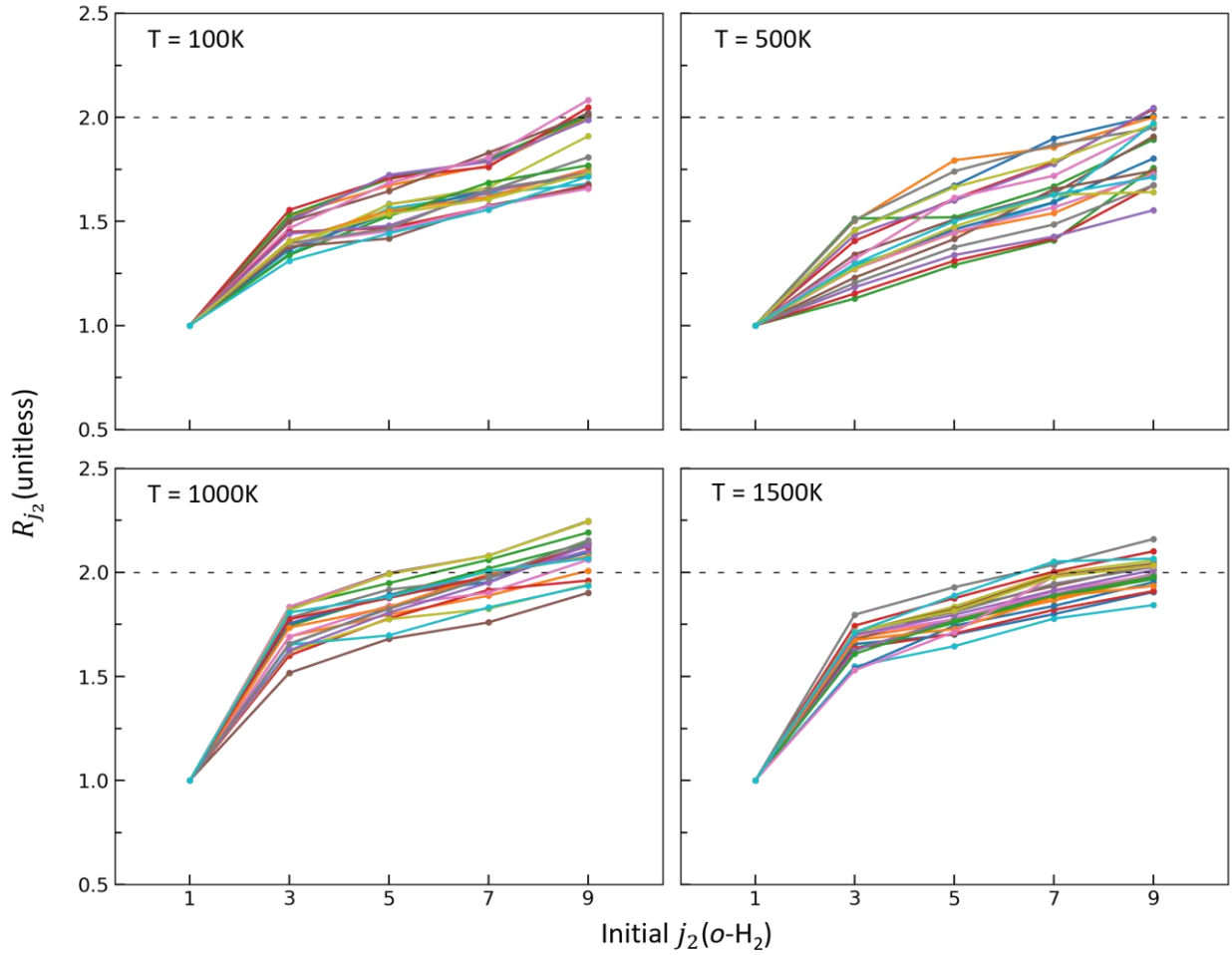

**Fig. S10:** Same as in Fig. S9, but for $p$-H$_2$O + $o$-H$_2$ projectile. The transitions are:

For T = 100K: $10_{010} \rightarrow 9_{19}$, $12_{57} \rightarrow 7_{17}$, $12_{48} \rightarrow 8_{08}$, $10_{55} \rightarrow 9_{64}$, $10_{46} \rightarrow 10_{37}$, $9_{46} \rightarrow 8_{53}$, $9_{19} \rightarrow 8_{08}$, $8_{62} \rightarrow 8_{53}$, $8_{53} \rightarrow 7_{44}$, $8_{44} \rightarrow 8_{35}$, $8_{44} \rightarrow 7_{35}$, $8_{44} \rightarrow 6_{42}$, $8_{26} \rightarrow 6_{24}$, $8_{08} \rightarrow 7_{17}$, $8_{08} \rightarrow 5_{15}$, $7_{71} \rightarrow 6_{60}$, $7_{62} \rightarrow 6_{51}$, $7_{53} \rightarrow 6_{42}$, $7_{44} \rightarrow 6_{33}$, and $7_{17} \rightarrow 6_{06}$.

For T = 500K: $10_{010} \rightarrow 9_{19}$, $12_{57} \rightarrow 12_{148}$, $10_{46} \rightarrow 10_{37}$, $9_{19} \rightarrow 8_{08}$, $8_{53} \rightarrow 7_{53}$, $8_{53} \rightarrow 7_{44}$, $8_{44} \rightarrow 7_{35}$, $8_{44} \rightarrow 6_{42}$, $8_{26} \rightarrow 6_{24}$, $8_{08} \rightarrow 7_{17}$, $7_{71} \rightarrow 6_{60}$, $7_{62} \rightarrow 6_{51}$, $7_{53} \rightarrow 6_{42}$, $7_{44} \rightarrow 6_{33}$, $7_{26} \rightarrow 5_{24}$, $7_{17} \rightarrow 6_{06}$, $6_{60} \rightarrow 5_{51}$, $6_{51} \rightarrow 5_{42}$, $6_{06} \rightarrow 5_{15}$, and $6_{42} \rightarrow 5_{33}$.

For T = 1000K: $10_{010} \rightarrow 9_{19}$, $12_{57} \rightarrow 12_{148}$, $8_{53} \rightarrow 7_{53}$, $8_{53} \rightarrow 7_{44}$, $8_{44} \rightarrow 7_{35}$, $8_{44} \rightarrow 7_{44}$, $8_{35} \rightarrow 6_{33}$, $8_{17} \rightarrow 7_{26}$, $8_{08} \rightarrow 7_{17}$, $7_{71} \rightarrow 6_{60}$, $7_{71} \rightarrow 5_{51}$, $7_{62} \rightarrow 6_{51}$, $7_{53} \rightarrow 7_{44}$, $7_{53} \rightarrow 6_{42}$, $7_{44} \rightarrow 7_{35}$, $7_{44} \rightarrow 6_{33}$, $7_{35} \rightarrow 6_{24}$, $7_{26} \rightarrow 6_{15}$, $7_{17} \rightarrow 6_{06}$, and $6_{60} \rightarrow 5_{51}$.

For T = 1500K: $11_{111} \rightarrow 10_{010}$, $11_{111} \rightarrow 9_{19}$, $10_{010} \rightarrow 9_{19}$, $10_{010} \rightarrow 8_{08}$, $10_{19} \rightarrow 9_{28}$ $10_{19} \rightarrow 8_{17}$, $9_{46} \rightarrow 8_{35}$, $9_{37} \rightarrow 8_{26}$, $9_{28} \rightarrow 8_{17}$, $9_{19} \rightarrow 8_{08}$, $8_{53} \rightarrow 7_{44}$, $8_{44} \rightarrow 7_{44}$, $8_{26} \rightarrow 8_{17}$, $8_{17} \rightarrow 7_{26}$, $8_{08} \rightarrow 7_{17}$ $7_{71} \rightarrow 6_{60}$, $7_{62} \rightarrow 6_{51}$, $7_{53} \rightarrow 6_{51}$, and $7_{53} \rightarrow 6_{42}$.

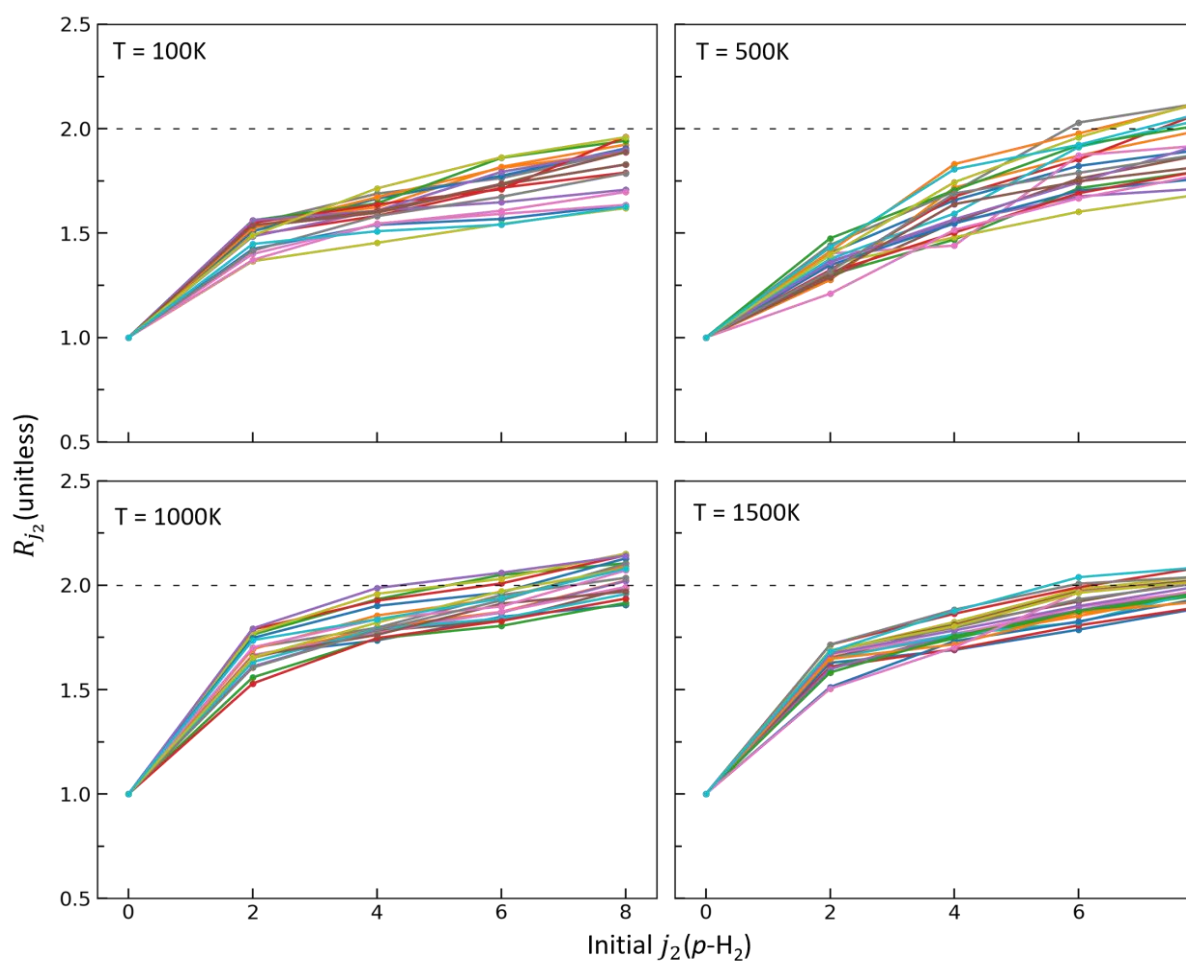

**Fig. S11:** The dependence of effective rate coefficients for $o$-$H_2O$ on the initial state $j_2$ of $p$-$H_2$ projectile, computed by MQCT for 20 most intense transitions for each individual temperature (indicated in the figure). The transitions are:

For T = 100K: $10_{010} \to 9_{09}$, $10_{110} \to 7_{07}$, $13_{49} \to 13_{310}$, $11_{83} \to 10_{92}$, $10_{74} \to 9_{81}$, $9_{54} \to 8_{45}$, $9_{45} \to 9_{36}$, $9_{09} \to 7_{07}$, $8_{54} \to 7_{43}$, $8_{18} \to 7_{07}$, $8_{18} \to 5_{05}$, $7_{70} \to 6_{61}$, $7_{61} \to 6_{52}$, $7_{52} \to 7_{43}$, $7_{52} \to 6_{43}$, $7_{43} \to 7_{34}$, $7_{16} \to 6_{25}$, $7_{07} \to 6_{16}$, $6_{61} \to 5_{50}$, and $6_{52} \to 6_{43}$.

For T = 500K: $10_{010} \to 9_{09}$, $10_{110} \to 8_{18}$, $12_{67} \to 12_{58}$, $11_{47} \to 11_{38}$, $10_{38} \to 10_{29}$, $9_{72} \to 9_{63}$, $9_{63} \to 9_{54}$, $9_{54} \to 9_{45}$, $9_{27} \to 7_{25}$, $9_{09} \to 8_{18}$, $8_{72} \to 8_{63}$, $8_{54} \to 7_{43}$, $8_{18} \to 7_{07}$, $7_{70} \to 6_{61}$, $7_{61} \to 6_{52}$, $7_{52} \to 6_{43}$, $7_{07} \to 6_{16}$, $6_{61} \to 5_{50}$, $6_{52} \to 5_{41}$, and $6_{16} \to 5_{05}$.

For T = 1000K: $10_{010} \to 9_{09}$, $10_{110} \to 8_{18}$, $12_{67} \to 12_{58}$, $11_{65} \to 10_{74}$, $11_{47} \to 11_{38}$, $10_{38} \to 10_{29}$, $9_{72} \to 7_{25}$, $9_{09} \to 8_{18}$, $8_{81} \to 8_{72}$, $8_{72} \to 8_{63}$, $8_{63} \to 8_{54}$, $8_{18} \to 7_{07}$, $7_{70} \to 6_{61}$, $7_{61} \to 6_{52}$, $7_{52} \to 6_{43}$, $7_{34} \to 7_{25}$, $7_{07} \to 6_{16}$, $6_{61} \to 5_{50}$, $6_{52} \to 5_{41}$, and $6_{16} \to 5_{05}$.

For T = 1500K: $10_{010} \to 9_{09}$, $10_{110} \to 9_{09}$, $12_{85} \to 11_{92}$, $11_{74} \to 11_{65}$, $11_{65} \to 11_{56}$, $11_{56} \to 11_{47}$, $10_{47} \to 10_{38}$, $9_{72} \to 9_{63}$, $9_{63} \to 9_{54}$, $9_{54} \to 9_{45}$, $9_{54} \to 8_{63}$, $9_{45} \to 9_{36}$, $9_{09} \to 8_{18}$, $8_{81} \to 8_{72}$, $8_{72} \to 8_{63}$, $8_{63} \to 8_{54}$, $8_{18} \to 7_{07}$, $7_{70} \to 6_{61}$, $7_{61} \to 6_{52}$, and $7_{07} \to 6_{16}$.

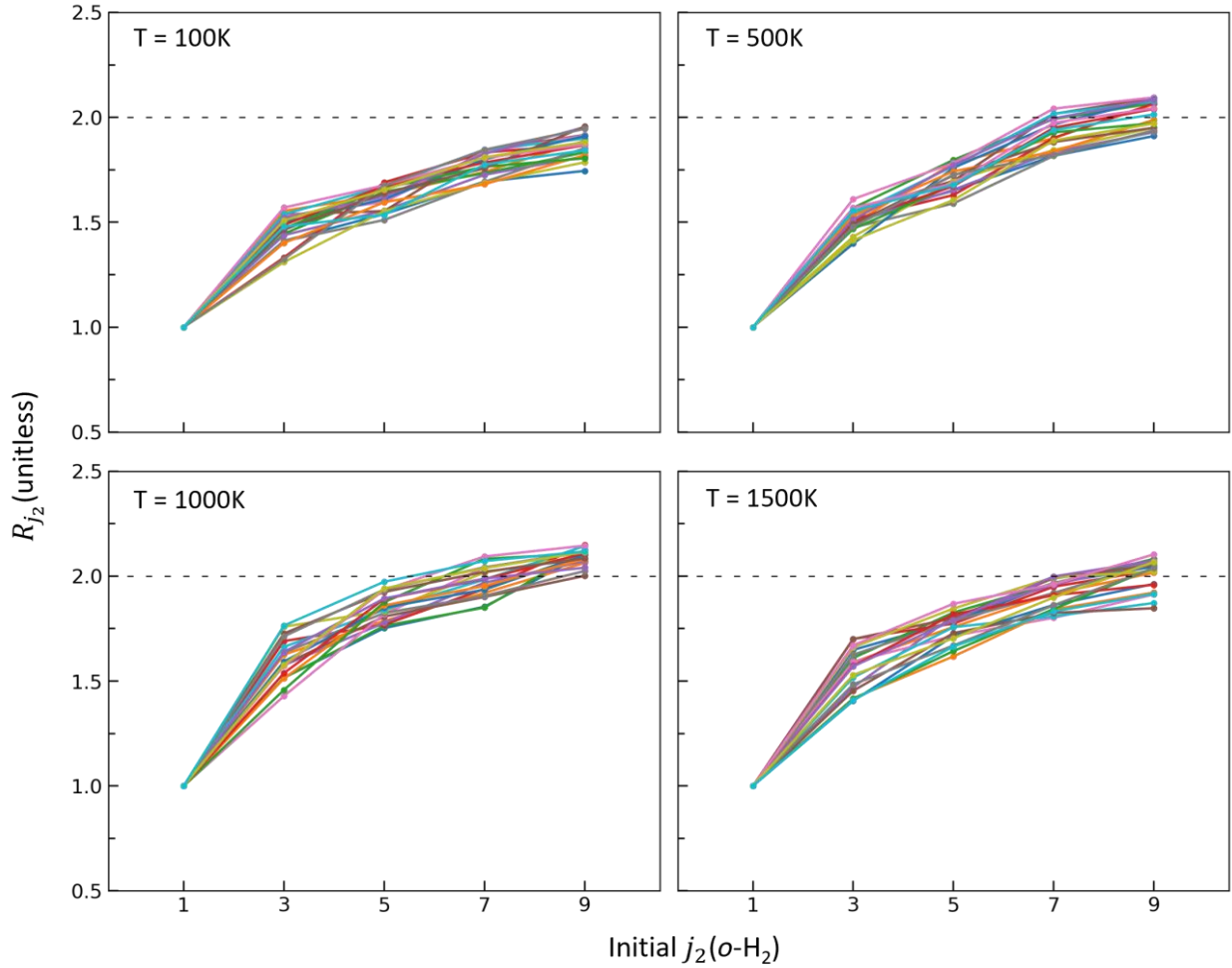

**Fig. S12:** Same as in Fig. S11, but for $o$-H$_2$O + $o$-H$_2$ projectile. The transitions are:

For T = 100K: $10_{010} \to 9_{09}$, $10_{110} \to 7_{07}$, $13_{49} \to 13_{310}$, $11_{83} \to 10_{92}$, $10_{74} \to 9_{81}$, $9_{54} \to 8_{45}$, $9_{45} \to 9_{36}$, $9_{09} \to 7_{07}$, $8_{54} \to 7_{43}$, $8_{18} \to 7_{07}$, $8_{18} \to 5_{05}$, $7_{70} \to 6_{61}$, $7_{61} \to 6_{52}$, $7_{52} \to 7_{43}$, $7_{52} \to 6_{43}$, $7_{43} \to 7_{34}$, $7_{16} \to 6_{25}$, $7_{07} \to 6_{16}$, $6_{61} \to 5_{50}$, and $6_{52} \to 6_{43}$.

For T = 500K: $10_{010} \to 9_{09}$, $10_{110} \to 8_{18}$, $12_{67} \to 12_{58}$, $11_{47} \to 11_{38}$, $10_{38} \to 10_{29}$, $9_{72} \to 9_{63}$, $9_{63} \to 9_{54}$, $9_{54} \to 9_{45}$, $9_{27} \to 7_{25}$, $9_{09} \to 8_{18}$, $8_{72} \to 8_{63}$, $8_{54} \to 7_{43}$, $8_{18} \to 7_{07}$, $7_{70} \to 6_{61}$, $7_{61} \to 6_{52}$, $7_{52} \to 6_{43}$, $7_{07} \to 6_{16}$, $6_{61} \to 5_{50}$, $6_{52} \to 5_{41}$, and $6_{16} \to 5_{05}$.

For T = 1000K: $10_{010} \to 9_{09}$, $10_{110} \to 8_{18}$, $12_{67} \to 12_{58}$, $11_{65} \to 10_{74}$, $11_{47} \to 11_{38}$, $10_{38} \to 10_{29}$, $9_{72} \to 7_{25}$, $9_{09} \to 8_{18}$, $8_{81} \to 8_{72}$, $8_{72} \to 8_{63}$, $8_{63} \to 8_{54}$, $8_{18} \to 7_{07}$, $7_{70} \to 6_{61}$, $7_{61} \to 6_{52}$, $7_{52} \to 6_{43}$, $7_{34} \to 7_{25}$, $7_{07} \to 6_{16}$, $6_{61} \to 5_{50}$, $6_{52} \to 5_{41}$, and $6_{16} \to 5_{05}$.

For T = 1500K: $10_{010} \to 9_{09}$, $10_{110} \to 9_{09}$, $12_{85} \to 11_{92}$, $11_{74} \to 11_{65}$, $11_{65} \to 11_{56}$, $11_{56} \to 11_{47}$, $10_{47} \to 10_{38}$, $9_{72} \to 9_{63}$, $9_{63} \to 9_{54}$, $9_{54} \to 9_{45}$, $9_{54} \to 8_{63}$, $9_{45} \to 9_{36}$, $9_{09} \to 8_{18}$, $8_{81} \to 8_{72}$, $8_{72} \to 8_{63}$, $8_{63} \to 8_{54}$, $8_{18} \to 7_{07}$, $7_{70} \to 6_{61}$, $7_{61} \to 6_{52}$, and $7_{07} \to 6_{16}$.